\documentstyle[12pt,a4,epsf,rotate]{article}
\newcommand{\rcite}[1]{{\cite{#1}}}
\newcommand{\rref}[1]{{(\ref{#1})}}
\newcommand{\tref}[1]{{\ref{#1}}}
\newcommand{\rlabel}[1]{{\label{#1}}}
\newcommand{\rbibitem}[1]{\bibitem{#1}}
\newcommand{\be}{\begin{equation}}
\newcommand{\ee}{\end{equation}}
\newcommand{\ba}{\begin{eqnarray}}
\newcommand{\ea}{\end{eqnarray}}
\newcommand{\dis}{\displaystyle}
\def\theequation{\arabic{section}.\arabic{equation}}
\newcommand{\mathrm}[1]{\mbox{\rm #1}}
\newcommand{\ovpi}{{\overline{\Pi}}}

\begin{document}
\begin{titlepage}
\begin{flushright}
NORDITA - 94/11 N,P
\end{flushright}
\vspace{2cm}
\begin{center}
{\Large\bf TWO- AND THREE-POINT FUNCTIONS IN THE EXTENDED
NJL MODEL}\\[2cm]
{\bf Johan Bijnens$^a$ and Joaquim Prades$^{a,b}$}\\[0.5cm]
${}^a$ NORDITA, Blegdamsvej 17,\\
DK-2100 Copenhagen \O, Denmark\\[0.5cm]
$^b$ Niels Bohr Institute, Blegdamsvej 17,\\
DK-2100 Copenhagen \O, Denmark\end{center}
\vfill
\begin{abstract}
The two-point functions in generalized Nambu--Jona-Lasinio
 models are calculated to all
orders in momenta and quark masses to leading order in $1/N_c$. The
use of Ward identities and the heat-kernel expansion allows for a large
degree of regularization independence. We also show 
how this approach works to the same order 
for three-point functions on the example
of the vector-pseudoscalar-pseudoscalar three-point function. 
The inclusion of the chiral anomaly effects
at this level is shown by calculating the pseudoscalar-vector-vector
three-point function to the same order. Finally we comment on how
(vector-)meson-dominance comes out in the presence of explicit chiral
symmetry breaking in both the anomalous and the non-anomalous sectors.
\end{abstract}
\vfill \begin{flushleft} March 1994\\revised April 1994
\end{flushleft} 
\end{titlepage}

\section{Introduction}
\rlabel{intro}

The Extended Nambu--Jona-Lasinio model (ENJL)\rcite{NJL,ENJL}
 has already a long history. For some
recent reviews see \rcite{review} and references therein. 
Generally a good agreement with low-energy
hadronic phenomenology has been found. However its main drawback
is the lack of confinement. In ref. \rcite{BBR} a large
number of relations between the observables was found which were valid in
a large class of ENJL-like models. In ref. \rcite{BRZ} this type of relations
was generalized to two-point functions and to all orders in the momenta
in the chiral limit.
Various numerical results obtained in ref. \rcite{Weise} thus obtained
a larger range of validity. In this work we shall extend this type of
analysis to three-point functions and two-point functions beyond
the chiral limit. The first one, the 
vector-pseudoscalar-pseudoscalar can already be found in \rcite{Lutz} and
we have included it to show explicitly the use of one-loop Ward identities
to simplify the calculation. A similar approach can be 
found in ref.\rcite{Meissner} which we  received 
 when the analytic part of this work was
essentially finished. They have a less general treatment of regularization
dependence than is done here and only treat the case with equal current-quark
masses. With the same definitions our results for the two-point functions
agree with theirs.
Our main aim is to apply this procedure to the case
of the pseudoscalar-vector-vector three-point function. 
Here we both illustrate our prescription for the consistent 
treatment of the chiral anomaly in this model by imposing the
QCD anomalous Ward identities \rcite{BP}. 
The latter do imply the use of consistent 
one-loop ENJL anomalous Ward identities.
We find that the 
duality between the Vector-Meson-Dominance (VMD) 
picture  for the slope of the anomalous $\pi^0\gamma
\gamma$ form factor and the quark-loop one
is much worse that the one  found for the pion 
electromagnetic form factor and a more refined
model (like ENJL cut-off like models) 
is necessary to reconcile both approaches.

At this point we would like to add some comments about the anomaly in
the ENJL model. In \rcite{Ball1} it has been argued that the anomaly can not
be consistently reproduced in this type of models. While we agree that
in general in these models no simple definition of the anomaly is possible,
we believe, as discussed in \rcite{BP}, that if one wants to use
these models as a low-energy approximation to QCD there is a 
unique prescription
of how to do this. Other uncertainties in the regularization scheme
of the anomalous sector will
be suppressed by powers of the cut-off $\Lambda_\chi$ 
used in the ENJL model.

The paper is organized as follows. 
In section \tref{second} we give a short
overview of the ENJL model.
Here we also discuss the dependence of the constituent quark mass on
the current quark mass and make some remarks about the definition of
quarks versus the QCD ones. In section \tref{twop} we extend the analysis
of ref. \rcite{BRZ} to the case with nonzero current quark masses and both
masses that play a r\^ole 
in the two-point function are allowed to be different.
We compare the results with Chiral Perturbation Theory 
($\chi$PT) and show numerical
results for some of the two-point functions. We also discuss the method used
shortly and give the new identities that the two-point functions at one-loop
and the full resummed ones have to satisfy. Their derivation is rather 
technical
and has been given explicitly in appendix \tref{AppB}. The main difference
with ref. \rcite{BRZ} is that now there is also non-trivial mixing in the
scalar sector. In subsection \tref{twop}.8 we discuss in detail the Weinberg
Sum Rules (WSR). 
It is found that here the high energy behaviour of this class
of ENJL-like models is too strongly suppressed.

Then we come to the derivation of the three-point functions here in the next
section \tref{threep}. In subsection \tref{vppsub} we discuss the 
vector-pseudoscalar-pseudoscalar three-point function paying attention to
the Ward identities in its calculation. In subsection \tref{pi0gg}
 we do the same for the pseudoscalar-vector-vector three-point 
function. Here we explain
how one needs to treat the anomalous part of the Ward identities to get
a consistent result.
Section \tref{VMD} treats the appearance of Vector Meson Dominance like
features of two- and three-point functions in this class 
of models. We briefly discuss two-point functions and particularly
the transverse
vector two-point function in the first subsection. Here the 
origin of the large shift in the slope compared to $M_V^2(0)$ of ref. 
\rcite{BBR} is explained. In subsection \tref{threevpp} we discuss the
VMD behaviour of the first three-point function and give a discussion
of the KSRF identity \rcite{KSRF} in this ENJL model.
 In the last subsection we treat the vector-pseudoscalar-pseudoscalar
three-point function
similarly. In section \tref{conclusions} we summarize our results.

The appendices contain the definition of our regularization procedure,
the derivation of the Ward identities and explicit expressions for the
one-loop functions we need.

\section{Short description of the ENJL model and its connection with QCD}
\rlabel{second}

The QCD Lagrangian is given by
\ba
\rlabel{QCD}
{\cal L}_{\rm QCD} &=& {\cal L}^0_{\rm QCD} -\frac{1}{4}G_{\mu\nu}
G^{\mu\nu} \, , \nonumber\\
{\cal L}^0_{\rm QCD} &=& \overline{q} \left\{i\gamma^\mu 
\left(\partial_\mu -i v_\mu -i a_\mu \gamma_5 - i G_\mu \right) - 
\left({\cal M} + s - i p \gamma_5 \right) \right\} q \, .
\nonumber \\
\ea
Here summation over colour degrees of freedom 
is understood and
we have used the following short-hand notations:
$\overline{q}\equiv\left( \overline{u},\overline{d},
\overline{s}\right)$; $G_\mu$ is the gluon field in the 
fundamental SU(N$_c$) (N$_c$=number
of colours) representation; 
$G_{\mu\nu}$ is the gluon field strength tensor in
the adjoint SU(N$_c$) representation; $v_\mu$, $a_\mu$, $s$ and
$p$ are external vector, axial-vector, scalar and pseudoscalar
field matrix sources; ${\cal M}$ is the quark-mass matrix. 

All indications are that in
the purely gluonic sector there is a mass-gap. Therefore there seems to be
a kind of cut-off mass in the gluon propagator (see the discussion in 
ref. \rcite{Lattice}). 
Alternatively one can think of integrating out the 
high-frequency (higher than $\Lambda_\chi$, a cut-off of the order 
of the spontaneous symmetry breaking scale) gluon
and quark degrees of freedom and then expand the resulting effective
action in terms of local fields. 
We then stop this expansion after the dimension
six terms. This leads to the following effective action
at leading order in the $1/N_c$ expansion 
\ba
\rlabel{ENJL}
{\cal L}_{\rm QCD} &\rightarrow& {\cal L}_{\rm QCD}^{\Lambda_\chi}
+ {\cal L}_{\rm NJL}^{\rm S,P} + {\cal L}_{\rm NJL}^{\rm V,A} +
{\cal O}\left(1/\Lambda_\chi^4\right),\nonumber\\ 
{\rm with}\hspace*{1.5cm} {\cal L}_{\rm NJL}^{\rm S,P}&=&
\frac{\dis 8\pi^2 G_S \left(\Lambda_\chi \right)}{\dis
N_c \Lambda_\chi^2} \, {\dis \sum_{i,j}} \left(\overline{q}^i_R
q^j_L\right) \left(\overline{q}^j_L q^i_R\right) \nonumber\\
{\rm and}\hspace*{3cm}&& \nonumber \\ 
{\cal L}_{\rm NJL}^{\rm V,P}&=&
-\frac{\dis 8\pi^2 G_V\left(\Lambda_\chi\right)}{\dis
N_c \Lambda_\chi^2}\, {\dis \sum_{i,j}} \left[
\left(\overline{q}^i_L \gamma^\mu q^j_L\right)
\left(\overline{q}^j_L \gamma_\mu q^i_L\right) + \left( L \rightarrow
R \right) \right] \,. \nonumber \\
\ea
Where $i,j$ are flavour indices and $\Psi_{R,L} \equiv
(1/2) \left(1 \pm \gamma_5\right) \Psi$. 
The couplings $G_S$ and $G_V$ are
dimensionless and ${\cal O}(1)$ in the $1/N_c$ expansion and summation 
over colours between brackets is understood. 
The Lagrangian ${\cal L}^{\Lambda_\chi}_{\rm QCD}$ includes only 
low-frequency modes of quark and gluon fields. The remaining
gluon fields can be assumed to be fully
absorbed in the coefficients of the local quark field operators 
or alternatively also described
by vacuum expectation values of gluonic operators (see the discussions 
in refs. \rcite{BBR,BRZ}). In the 
mean-field approximation these ${\cal L}_{\rm NJL}^{\rm S,P,V,A}$ above
are equivalent to a constituent chiral quark-mass term \rcite{ERT}.

This model has the same symmetry structure as the QCD action 
at leading order in $1/N_c$ \rcite{tHooft}
(notice that the $U(1)_A$
problem is absent at this order \rcite{Witten}).
(For explicit symmetry properties under SU(3)$_L$ $\times$
SU(3)$_R$ of the fields in this model
see reference \rcite{BBR}.)
We can self-consistently solve the Schwinger-Dyson equation for
the fermion propagator in terms of the bare propagator and a 
one-loop diagram.
In the case where the current quark masses are set to 
zero this equation allows for
two solutions for $G_S > 1$, one with constituent quark mass
$M = 0$ and the other with  $M\ne 0$
and the model shows spontaneous chiral symmetry breaking.
In the presence of explicit chiral symmetry breaking only the second solution
is allowed. We shall allow for nonzero current quark masses, ${\cal M}=
{\rm diag} \left(m_u,m_d,m_s\right)$ and
all different. In the leading $1/N_c$ limit the solution of the Schwinger-Dyson
equation is a flavour diagonal matrix for the constituent quark masses with 
elements $M_{u,d,s}$.
The gap-equation now becomes
\ba
\rlabel{gap}
M_i &=& m_i - g_S \langle 0| :\overline{q}_i q_i : | 0\rangle \, ,
\\
\rlabel{VEV}
\langle 0 | : \overline{q}_i q_i : | 0 \rangle \equiv
\langle  \overline{q}_i q_i \rangle &=& - N_c 4 M_i 
\int \frac{{\rm d}^4p}{(2\pi)^4}
\frac{i}{p^2-M_i^2} \nonumber\\
&=& - \frac{\dis N_c}{\dis 16 \pi^2} 
4 M_i^3  \Gamma\left(-1, \epsilon_i\right) \, ,
\\ \rlabel{gs}
g_S &\equiv& \frac{4\pi^2 G_S}{N_c\Lambda_\chi^2}~.
\ea

Therefore, in this model the scalar quark-antiquark one-point function 
(quark condensate)  obtains a non-trivial nonzero value.
The dependence on the current quark-mass is somewhat obscured
 in eq. \rref{VEV}.
We use here a cut-off in proper time as the regulator. See appendix 
\ref{GAMMA} for its definition. The quantity $\epsilon_i$ appearing in 
\rref{VEV} is $M_i^2/\Lambda_\chi^2$. In figure \ref{Figvev} we have plotted
the dependence of $M_i$ on $G_S$ for various values of $m_i$
and $\Lambda_\chi = 1.160$ GeV.
\begin{figure}
\rotate[r]{\epsfysize=13.5cm\epsfxsize=8cm\epsfbox{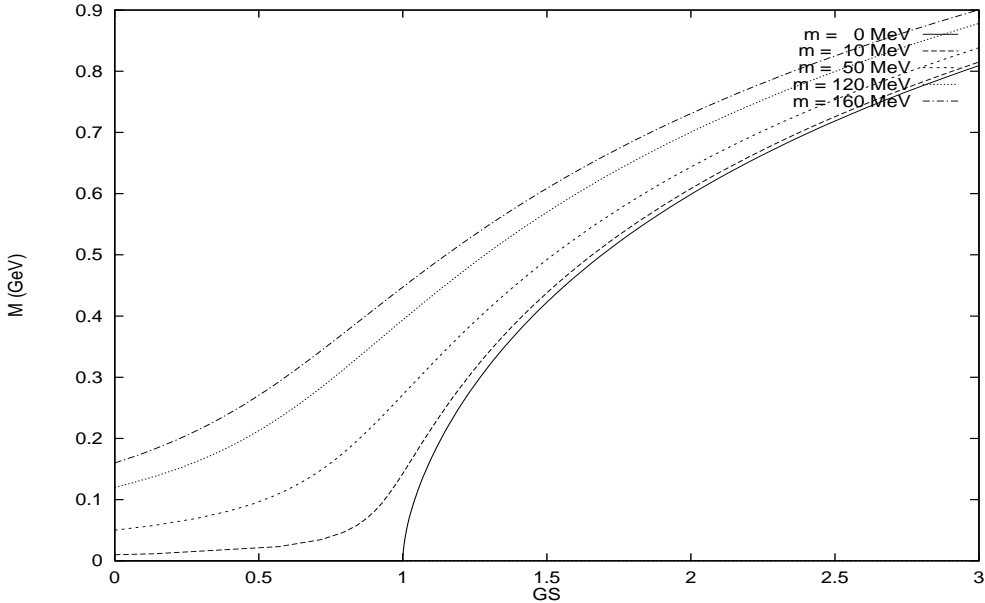}}
\caption{Plot of the dependence of the constituent quark mass $M_i$ as a
function of $G_S$ for several values of $m_i$}
\rlabel{Figvev}
\end{figure}
It can be seen that the value of $M_i$ for small $m_i$ converges smoothly
towards the value in the chiral limit for the spontaneously broken phase. 
This is an indication that an expansion in the quark masses 
as Chiral Perturbation Theory assumes for QCD is also valid in this model. 
However, it can also be seen that the validity of this expansion
breaks down quickly and for $m_i \simeq 200$ MeV we already
have $2 M_i \simeq \Lambda_\chi$.
We note that the ratio of vacuum expectation values
 for light quark flavours increases with increasing 
current quark mass 
at $p^2=0$ in this model and starts to saturate
 for $m_i > 200$ MeV. In standard $\chi$PT this ratio is taken
to be 1 at lowest order 
and its  behaviour with the current quark mass
is governed (at ${\cal O} (p^4)$) by the following combination
of coupling constants $2 L_8 + H_2$ \rcite{GL} in the
large $N_c$ limit. The 
${\cal O} (p^4)$ $\chi$PT coupling constants \rcite{GL} are calculated
at leading order in $1/N_c$ and in the chiral limit 
in the ENJL model
\rcite{BBR}\footnote{The analytical expression for $H_2$ in that 
reference is correct. The tables contain a numerical error. For example
the value of $H_2$ for the parameters of fit 1 in 
ref. \rcite{BBR} is  $1.4 \cdot 10^{-3}$}.
The analytical result for this combination of couplings constants
there was confirmed in ref. \rcite{BRZ} from a calculation of the
scalar two-point function in the chiral limit and in the large
$N_c$ limit to all orders in momenta. This result was
obtained by requiring the relevant Ward identities to all orders
in momenta.
The value for the combination $2 L_8 + H_2$ found there corresponds
to a big increase of the quark condensate with the current quark mass
at this order. We want to emphasize here that
the exact identification of the quark condensate in eq. \rref{VEV} 
which is regularization dependent and especially its dependence on the
current quark mass with the one used in $\chi$PT or QCD Sum
Rules is by no means straightforward. The differences between both 
can be traced back in the scalar sector of the model and 
in particular in the quadratically regularization dependent 
${\cal O} (p^4)$ coupling constant $H_2$ \rcite{GL}.
This high-energy constant is related to the details of the integration 
of the  QCD high-frequency modes to obtain the Lagrangian in eq. 
\rref{ENJL}.  
However there are indications that the QCD light quark condensates
indeed increase with the current quark mass.
In general, there is some uncertainty in the definition of the quark
fields in ENJL models versus the QCD ones. This depends on the details
on how the ENJL model originates from QCD.

\setcounter{equation}{0}
\section{Two-point functions in the presence 
of current quark masses}
\rlabel{twop}

This section is a generalization of the results in ref. \rcite{BRZ} to
the case of nonzero current quark masses. These two-point functions were
studied before in \rcite{Weise} but there they were discussed as quark
form factors. What is new here is that the explicit dependence on
the regularization scheme has been put into two arbitrary functions,
namely, $\ovpi_V^{(0)}+\ovpi_V^{(1)}$ and $\ovpi^P_M$
(see this section below for definitions). This
also shows that these results are valid in a class of models where the
one-loop (see further for the definition of this) result can be expanded
in a heat-kernel expansion using the same basic quantities $E$ and $R_{\mu\nu}$
as used here. This includes the ENJL model with low-energy gluons described
by background expectation values. We have not included this case in our 
numerical results for the explicit one-loop expressions. For the
equal mass case the relevant one loop formulas can be 
found in ref. \rcite{BRZ}.

\subsection{Definition of the two-point functions}

We shall discuss two--point functions of the vector,
axial--vector, scalar and pseudoscalar quark currents
with the following definitions, 
\begin{eqnarray}
\rlabel{1}
V^{ij}_\mu (x)&\equiv& \bar q_i(x)\gamma_\mu
q_j(x)\, ,
\\
\rlabel{2}
A^{ij}_\mu (x)&\equiv& \bar q_i(x)\gamma_\mu\gamma_5
q_j(x)\, ,
\\
\rlabel{3}
S^{ij} (x)&\equiv& -\, \bar q_i(x) q_j(x)\, ,
\\
\rlabel{4}
P^{ij} (x)&\equiv& \bar q_i(x)\, i\gamma_5 q_j(x) \, ,
~.
\end{eqnarray}
The indices $i,j$ are flavour indices and run over $u,d,s$. The two-point
functions themselves are defined as
\ba
\rlabel{9}
{ \Pi}^V_{\mu\nu} (q)_{ijkl}&=& i\int {\rm d}^4x e^{iq\cdot x}
<0|T\left( V_\mu^{ij} (x)V_\nu^{kl}(0)\right) |0> \, ,
\\
\rlabel{10}
{ \Pi}^A_{\mu\nu} (q)_{ijkl}&=& i\int {\rm d}^4x e^{iq\cdot x}
<0|T\left( A_\mu^{ij} (x)A_\nu^{kl}(0)\right) |0>\, ,
\\
\rlabel{11}
{ \Pi}^S_{\mu} (q)_{ijkl}&=& i\int {\rm d}^4x e^{iq\cdot x}
<0|T\left( V_\mu^{ij} (x)S^{kl}(0)\right) |0>\, ,
\\
\rlabel{12}
{     \Pi}^P_{\mu} (q)_{ijkl}&=& i\int {\rm d}^4x e^{iq\cdot x}
<0|T\left( A_\mu^{ij} (x)P^{kl}(0)\right) |0>\, ,
\\
\rlabel{13}
{     \Pi}^S (q)_{ijkl}&=& i\int {\rm d}^4x e^{iq\cdot x}
<0|T\left( S^{ij} (x)S^{kl}(0)\right) |0>\, ,
\\
\rlabel{14}
{     \Pi}^P (q)_{ijkl}&=& i\int {\rm d}^4x e^{iq\cdot x}
<0|T\left( P^{ij} (x)P^{kl}(0)\right) |0>\ .
\ea
In the leading order in the number of colours these are all 
proportional to $\delta_{ijkl} \equiv
 \delta_{il}\delta_{jk}$, with $\delta_{il}$
the Kronecker delta. Using Lorentz-invariance these functions can then
be expressed as follows
\ba
\rlabel{15}
{     \Pi^V_{\mu\nu} (q)_{ijkl}}&=& \left\{
(q_\mu q_\nu -q^2g_{\mu\nu}){ \Pi_V^{(1)}}(Q^2)_{ij} + 
q_\mu q_\nu {\Pi_V^{(0)}}(Q^2)_{ij} \right\}
\delta_{ijkl} \, , 
\\
\rlabel{16}
{\Pi^A_{\mu\nu} (q)_{ijkl}}&=& \left\{ (q_\mu q_\nu -q^2g_{\mu\nu}){  
   \Pi_A^{(1)}}(Q^2)_{ij} + q_\mu q_\nu {\Pi_A^{(0)}}(Q^2)_{ij} 
\right\} \delta_{ijkl} \, ,
\\
\rlabel{17}
{     \Pi^S_{\mu} (q)_{ijkl}}&=& q_\mu \, {     \Pi^M_
{S}}(Q^2)_{ij} \delta_{ijkl} \, ,
\\
\rlabel{18}
{     \Pi^P_{\mu} (q)_{ijkl}}&=& i q_\mu \,  
{     \Pi^M_{P}}(Q^2)_{ij} \delta_{ijkl} \, ,
\\
\rlabel{19}
{     \Pi^S(q)_{ijkl}}&=& {\Pi_{S}}(Q^2)_{ij}\delta_{ijkl} \, ,
\\
\rlabel{20}
{     \Pi^P(q)_{ijkl}}&=& {     \Pi_
{P}}(Q^2)_{ij}\delta_{ijkl}\ .
\ea
Here $Q^2 = -q^2$. We shall discuss the Weinberg Sum Rules 
and numerical results
for the two-point functions only in the Euclidean domain,
i.e. $Q^2$ positive.
Using Bose symmetry on the definitions of the two-point functions 
it follows
that $\Pi^{(0)}_{V}(Q^2)_{ij}$,
$\Pi^{(1)}_{V}(Q^2)_{ij}$,
$\Pi^{(0)}_{A}(Q^2)_{ij}$,
$\Pi^{(1)}_{A}(Q^2)_{ij}$,
$\Pi_{S}(Q^2)_{ij}$ and
$\Pi_{M}(Q^2)_{ij}$ are all symmetric in the flavour indices $i$ and $j$.
The remaining ones need the Ward-identities to prove 
their flavour structure. From the 
identities in the appendix \ref{AppB} 
it follows that $\Pi^M_S(Q^2)_{ij}$
is also symmetric in $i,j$; while  
$\Pi^M_S(Q^2)_{ij}$ is anti-symmetric. 

\subsection{Lowest order results in Chiral 
Perturbation Theory} From Chiral Perturbation 
Theory to order $p^4$ in the expansion 
we obtain the following
low energy results for the two-point functions. The orders mentioned behind
are the orders in Chiral Perturbation Theory that are neglected. 
\ba
\rlabel{21}
{    \Pi_V^{(1)}}(Q^2)_{ij} &=& -4(2H_1 + L_{10}) + {\cal O}(p^6) \, ,
\\
\rlabel{22}
{    \Pi_V^{(0)}}(Q^2)_{ij} &=& {\cal O}(p^6)\, ,
\\
\rlabel{23}
{\Pi_A^{(1)}}(Q^2)_{ij}&=& {2f_{ij}^2\over Q^2}-4(2H_1 - L_{10}) + 
{\cal O}(p^6)\, ,
\\
\rlabel{24}
{\Pi_A^{(0)}}(Q^2)_{ij}&=& 2f_{ij}^2 \left(\frac{1}{m_{ij}^2
+ Q^2} - \frac{1}{Q^2}\right) + {\cal O} (p^6)\, ,
\\
\rlabel{25}
{    \Pi^M_S    }(Q^2)_{ij}&=& {\cal O}(p^6) \, ,
\\
\rlabel{26}
{    \Pi^M_P    }(Q^2)_{ij}&=& {2B_0f_{ij}^2\over 
m_{ij}^2+Q^2}+{\cal O}(p^6) \, ,
\\
\rlabel{27}
{    \Pi_S    }(Q^2)_{ij}&=& 8B_0^2(2L_8 + H_2) + {\cal O}(p^6) \, ,
\\
\rlabel{28}
{\Pi}_P (Q^2)_{ij}&=& {\dis 2B_0^2 f_{ij}^2 \over  m_{ij}^2 + Q^2} 
+ 8B_0^2(-2L_8+H_2) + {\cal O} (p^6)\ .
\ea
With $m_{ij}$ the mass of the lightest pseudoscalar meson with flavour
structure $ij$.
These are obtained in the leading $1/N_c$ approximation so loop-effects are
not needed. Notice that these expressions 
are valid to chiral order $p^4$. From a term of the 
form $\mathrm{tr}\{D_\mu\chi D^\mu\chi^\dagger\}$ there
are contributions of order $(m_i-m_j)^2/Q^2$  to the vector two-point function
$\Pi^{(0)}_V(Q^2)_{ij}$
and of order $(m_i-m_j)$ to the mixed scalar vector function
$\Pi^M_S(Q^2)_{ij}$.

The functions ${\Pi_A^{(0)}}$, ${\Pi^M_P}$ and ${
\Pi_P}$ 
get their leading behaviour from the pseudoscalar Goldstone pole.
In addition $\Pi_A^{(1)}$ and $\Pi_A^{(0)}$ contain a kinematical
pole at $Q^2=0$.
The residue of the physical 
pole is proportional to the decay constant $f_{ij}$
for the relevant meson,
(for the $\bar u d$ ones, $f_{ud}\simeq f_\pi \simeq 92.5$ MeV).
In $\chi$PT, the constant $B_0$ is related to the vacuum
expectation value in the chiral limit. 
In the large $N_c$ limit and away from the chiral limit there are
corrections due to the terms proportional to combination of 
${\cal O}(p^4)$ couplings $2L_8+H_2$ \rcite{GL}. 
\be\rlabel{29}
<0|:{\overline \Psi} \Psi :|0>_{\left|\Psi=u,d,s\right.}\equiv
-f_0^2 B_0\, \left ( 1+{\cal O}(p^4)\right)\ .
\ee
The vacuum expectation value here, $<0|: {\overline \Psi} \Psi : |0>$,
is the one used in $\chi$PT in the chiral limit and $f_0$ is the
pseudoscalar meson decay constant in the chiral limit.
The constants $L_8$, $L_{10}$, $H_1$ and $H_2$ are coupling constants
of the ${\cal O}(p^4)$ effective chiral Lagrangian in the notation of
Gasser and Leutwyler \rcite{GL}. The constants $L_8$ and $L_{10}$ 
are known from the comparison 
between $\chi$PT and low energy hadron phenomenology.
At the scale of the $\rho$ meson mass they are
$L_8=(0.9\pm 0.3)\times 10^{-3}$  and
$L_{10}=(-5.5\pm 0.7)\times 10^{-3}$.
The high energy constants
 $H_1$ and $H_2$ correspond to couplings which
involve external source fields only and therefore can only 
be extracted from experiment given a prescription.

\subsection{The method and Ward identities}

The method used here is identical to the one used in \rcite{BRZ}.
The full two-point functions are the sum of diagrams like those in 
figure \ref{Fig2pt}a.
The one-loop two-point functions are those obtained by the graph in
figure \ref{Fig2pt}b. 
\begin{figure}
\begin{center}
%
%
%
\thicklines
\setlength{\unitlength}{1mm}
\begin{picture}(140.00,50.00)
\put(97.50,35.00){\oval(15.00,10.00)}
\put(103.00,33.50){$\bigotimes$}
\put(17.50,35.00){\oval(15.00,10.00)}
\put(25.00,35.00){\circle*{2.00}}
\put(32.50,35.00){\oval(15.00,10.00)}
\put(40.00,35.00){\circle*{2.00}}
\put(47.50,35.00){\oval(15.00,10.00)}
\put(55.00,35.00){\circle*{2.00}}
\put(62.50,35.00){\oval(15.00,10.00)}
\put(08.00,33.50){$\bigotimes$}
\put(68.00,33.50){$\bigotimes$}
\put(88.00,33.50){$\bigotimes$}
\put(38.50,19.00){(a)}
\put(95.50,19.00){(b)}
\put(14.50,40.00){\vector(1,0){3.00}}
\put(29.50,40.00){\vector(1,0){3.50}}
\put(44.00,40.00){\vector(1,0){5.00}}
\put(60.50,40.00){\vector(1,0){3.00}}
\put(95.50,40.00){\vector(1,0){5.00}}
\put(99.00,30.00){\vector(-1,0){3.00}}
\put(64.00,30.00){\vector(-1,0){3.00}}
\put(49.50,30.00){\vector(-1,0){3.50}}
\put(34.00,30.00){\vector(-4,1){2.00}}
\put(18.00,30.00){\vector(-1,0){2.50}}
\end{picture}
\caption{The graphs contributing to the two point-functions
in the large $N_c$ limit. 
a) The class of all strings of constituent quark loops.
The four-fermion vertices are either  
${\cal L}^{\rm S,P}_{\rm NJL}$ or  
${\cal L}^{\rm V,A}_{\rm NJL}$ in eq. \protect{\rref{ENJL}}.
The crosses at both ends are the insertion of the external sources. 
b) The one-loop case.}
\rlabel{Fig2pt}
\end{center}
\end{figure}
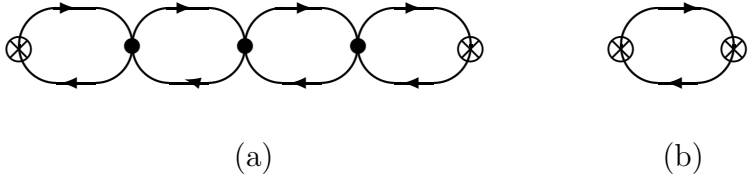
Using a recursion formula that relates the
$n$-loop graph to a product of the one-loop and the ($n$--1)-loop graph
and the relevant combination of kinematic factors and $G_V$ and $G_S$ the
whole class of graphs can be easily summed. Some care must be taken in the
case where different two-point functions can mix so a matrix
inversion is necessary (see ref. \rcite{BRZ}).

The two-point functions defined above satisfy the following 
Ward identities.
(We suppress the argument $Q^2$ for brevity.)
\ba
\rlabel{ward1}
-Q^2{\Pi^{(0)}_V}_{ij} &=& \left( m_i - m_j \right){\Pi_S^M}_{ij} \, ,
\\
\rlabel{ward2}
-Q^2{\Pi_S^M}_{ij} &=& \left( m_i - m_j \right){\Pi_S}_{ij}
+\langle \overline{q}_i q_i\rangle - \langle \overline{q}_j q_j
\rangle \, ,
\\
\rlabel{ward3}
-Q^2{\Pi^{(0)}_A}_{ij} &=& \left( m_i + m_j \right){\Pi_P^M}_{ij} \, ,
\\
\rlabel{ward4}
-Q^2{\Pi_P^M}_{ij} &=& \left( m_i + m_j \right){\Pi_P}_{ij}
+\langle \overline{q}_i q_i\rangle + \langle \overline{q}_j q_j
\rangle \, .
\ea
These are derived in the appendix \ref{AppB}. From these the 
flavour symmetry of
the mixed two-point functions can be derived from the vector ones.

The one-loop expressions, which we shall 
denote by $\overline{\Pi}$ and use
further the same conventions as given for the full ones above
are given in appendix \ref{AppC}. They
satisfy the same identities but with the current quark masses $m_i$ replaced
by the constituent ones, $M_i$. In addition to these, there are two
more relations that follow in general if the one-loop part can be 
described by a heat-kernel expansion in terms of the quantities $E$ and
$R_{\mu\nu}$ of appendix \ref{AppB}.
These identities are (with the flavour subscript $ij$ and argument 
suppressed)
\ba
\rlabel{ward5}
\overline{\Pi}^{(1)}_V +  \overline{\Pi}^{(0)}_V &=&
\overline{\Pi}^{(1)}_A +  \overline{\Pi}^{(0)}_A \, ,
\\ \rlabel{ward6}
\overline{\Pi}_S + Q^2 \overline{\Pi}^{(0)}_V &=&
\overline{\Pi}_P + Q^2 \overline{\Pi}^{(0)}_A \, .
\ea

\subsection{The transverse vector sector}
\rlabel{vector1}
We introduce here for convenience an extra symbol $g_V$
\be
\rlabel{GV}
g_V \equiv \frac{8\pi^2G_V}{N_c\Lambda_\chi^2} Q^2\ .
\ee
The full resummed transverse vector two-point function is then
\be
\rlabel{vector}
{\Pi^{(1)}_V}_{ij} = 
\frac{\overline{\Pi}^{(1)}_{Vij}}
{1 + g_V \overline{\Pi}^{(1)}_{Vij}}\ .
\ee
This can be simply written in a form resembling the
one in the complete VMD limit 
with couplings $f_S$, $f_V$ and $M_V$ depending on 
$Q^2$ and flavour and defined by
\ba
\rlabel{transvec}
\Pi^{(1)}_V(Q^2)_{ij} &=& \frac{2f_S^2(Q^2)_{ij}}{Q^2} + 
\frac{2 f_V^2(Q^2)_{ij} M_V^2(Q^2)_{ij}}{M_V^2(Q^2)_{ij} + Q^2} \, ,
\\
2f_S^2(Q^2)_{ij}&=& \frac{-Q^2\ovpi^{(0)}_V(Q^2)_{ij}}
{1-g_V \ovpi^{(0)}_V(Q^2)_{ij}} \, ,
\\
\rlabel{fvmv}
2 f_V^2(Q^2)_{ij} M_V^2(Q^2)_{ij} &=& \frac{N_c\Lambda_\chi^2}
{8\pi^2 G_V} \frac{1}{1-g_V \ovpi^{(0)}_V(Q^2)_{ij}} \, ,
\\
2 f_V^2(Q^2)_{ij} &=& {\overline{\Pi}^{(0+1)}_V(Q^2)_{ij}}\ .
\ea
Where we have used the fact that (see appendices
 \ref{AppB} and \ref{AppC})
$\ovpi^{(0+1)}_V \equiv \ovpi^{(0)}_V + \ovpi_V^{(1)}$ 
has no pole at $Q^2 = 0$.
There is a correction here (in $\ovpi^{(0)}_V$)
due to the mixing with the scalar sector, which is allowed
by  the presence of explicit breaking of the vector symmetry
(see the scalar mixed sector subsection \tref{scalar0}).
 For the diagonal case, this is defined
as $m_i = m_j$ or $M_i = M_j$,  
$\overline{\Pi}^{(0)}_V$ vanishes and the formulas above
simplify very much. 

The pole mass of the vector corresponds to the pole in 
this two point function
or to the solution of 
$\mathrm{Re}\left( Q^2 + M_V^2(Q^2)_{ij} \right) = 0$.
Alternatively, one can define the VMD values for the vector
parameters ($f_V$ and $M_V$) 
as the best parameters of a linear fit of 
the inverse of $\Pi^{(1)}_V(Q^2)_{ij}-2f^2_S(Q^2)_{ij}/Q^2$. 
These definitions have the advantage
that they are also valid for the Euclidean region ($Q^2>0$)
where the vector cannot decay into two constituent
quarks. See sections
on numerical applications \tref{numbers} and 
Vector-Meson-Dominance  \tref{VMD} for further comments.

\subsection{The transverse axial-vector sector}
\rlabel{axialvector1}
The transverse 
axial-vector two-point function derivation is also identical to
the one in ref. \rcite{BRZ}.
\be
\rlabel{axvector}
{\Pi^{(1)}_A}_{ij} = 
\frac{\overline{\Pi}^{(1)}_{Aij}}{1 + g_V \overline{\Pi}^{(1)}_{Aij}}\ .
\ee
Using the identity \rref{ward5} 
it can be seen that this has a pole
at $Q^2$=0 because $\ovpi^{(0)}_{A}$ has it. As can be seen from
the explicit expression and is proved in general 
in appendix \ref{AppB}, the
combination $\overline{\Pi}^{(0)}_V + \overline{\Pi}^{(1)}_V$ is regular
at $Q^2$ going to zero.  This again allows us 
to separate the pole at $Q^2=0$ in a simple fashion.
\ba \rlabel{46}
 {\Pi}_A^{(1)}(Q^2)_{ij}
& =&  {2f_{ij}^2(Q^2)\over Q^2}
 + {2f_A^2(Q^2)_{ij} M_A^2(Q^2)_{ij}\over M_A^2(Q^2)_{ij}+Q^2}\ ,
\\ \rlabel{104}
f_{ij}^2(Q^2) &=& g_A(Q^2)_{ij}\bar f_{ij}^2(Q^2) \ ,
\\ \rlabel{105}
2\bar f_{ij}^2(Q^2)&=&-Q^2 \overline{\Pi}^{(0)}_A(Q^2)_{ij}\ ,
\\
\left(g_A(Q^2)_{ij}\right)^{-1} &=&
1- g_V \overline{\Pi}^{(0)}_A(Q^2)_{ij}\ ,
\\
\rlabel{fa}
2 f_A^2(Q^2)_{ij} M_A^2(Q^2)_{ij} &=&
\frac{N_c\Lambda_\chi^2}{8\pi^2 G_V} g_A(Q^2)_{ij} \, ,
\\
2 f_A^2(Q^2)_{ij} &=& g_A^2(Q^2)_{ij} 
\ovpi^{(0+1)}_V(Q^2)_{ij}\ .
\ea
There is a correction here (in $\ovpi^{(0)}_A$)
due to the mixing with the pseudo-scalar sector due to
the presence of both spontaneous and 
explicit breaking of the axial-vector symmetry
(see the pseudo-scalar mixed sector subsection).
For further
 discussion of these expressions and the ones in the previous 
section we refer to the subsection \tref{WSRsec} on
Weinberg Sum Rules.

\subsection{The pseudo-scalar mixed sector}
\rlabel{pseudo0}

The same method as used in \rcite{BRZ} still applies with the results for
the summed functions given in terms of the function 
$\Delta_P(Q^2)$ and the one loop two-point functions 
(with flavour subscripts $ij$ suppressed),
\ba
\rlabel{131}
 {    \Pi}_A^{(0)}(Q^2)&=& {1\over \Delta_P(Q^2)} \left[
(1-g_S \overline{ \Pi}_P(Q^2))\overline{\Pi}_A^{(0)}(Q^2) 
+ g_S{(\overline{\Pi}_P^M(Q^2))}^2
\right] \ ,
\\
\rlabel{132}
{\Pi}_P^M(Q^2)&=&{1\over \Delta_P (Q^2)}  
{\overline\Pi}_P^M(Q^2)\ ,
\\
\rlabel{133}
 {    \Pi}_P(Q^2)&=& {1\over \Delta_P (Q^2)} \left[
(1-g_V\overline{    \Pi}_A^{(0)}(Q^2))\overline{\Pi}_P (Q^2)+
 g_V{(\overline{    \Pi}^M_P(Q^2))}^2 \right] \, ,\\
\rlabel{134}
\Delta_P (Q^2)&=&\left( 1-g_V\overline{    \Pi}_A^{(0)}
(Q^2)\right)
 \left( 1 -g_S\overline{    \Pi
}_P(Q^2)\right)   - g_Sg_V{\left( \overline{    \Pi}^M_P
(Q^2)\right) }^2 .
\ea
Using the identities for the one-loop case it can be shown that the resummed 
ones satisfy the Ward identities of appendix \ref{AppB} with the
current quark masses. To show this it is also necessary to use the
Schwinger-Dyson equation for the constituent quark masses in 
eq.\rref{gap}.
 
In order to rewrite this in terms of a nicer notation 
we first express $\Delta_P(Q^2)_{ij}$ 
in a different form using the identities for the one-loop
two-point functions.
\ba
\Delta_P(Q^2)_{ij} &=& \frac{g_S \ovpi^M_P (Q^2)_{ij}}{M_i+M_j}
\left( m_{ij}^2(Q^2) + Q^2\right) 
\\
\rlabel{mpi}
{\rm with } \hspace*{0.2cm} 
m_{ij}^2(Q^2) &\equiv& \frac{\left(m_i+m_j\right)}
{g_S g_A(Q^2) \ovpi^M_P(Q^2)_{ij}}\ .
\ea
Inserting the definition of $f_{ij}^2(Q^2)$ and
$1/g_S = -\langle \overline{q}_iq_i\rangle/\left(M_i-m_i\right)$ we
recover the Gell-Mann--Oakes--Renner (GMOR) relation for the pion mass 
\rcite{Gell} when eq. \rref{mpi} is expanded in powers of $m_i$. 
For further discussion on corrections to the GMOR relation
in this model we refer to the section on numerical applications
\tref{numbers}.
Formula \rref{mpi} gives
the expression for the pole due to the lightest pseudoscalar
mesons in the presence of explicit
chiral symmetry breaking.

This then allows us to 
rewrite the full two-point functions in a very simple fashion:
\ba
\rlabel{pipa0}
\Pi^{(0)}_A(Q^2)_{ij} &=& 2 f_{ij}^2(Q^2)
\left(\frac{1}{m_{ij}^2(Q^2)+Q^2}
 - \frac{1}{Q^2}\right) \, ,
\\ \rlabel{pipm}
\Pi_P^M(Q^2)_{ij} &=& \frac{M_i + M_j}{g_S}\frac{1}
{m_{ij}^2(Q^2)+Q^2} \, ,
\\ \rlabel{pip}
\Pi_P(Q^2)_{ij} &=& -\frac{1}{g_S} + \frac{\left(M_i+M_j\right)^2}
{2 f_{ij}^2(Q^2)} \frac{1}{g_S^2}\frac{1}
{m_{ij}^2(Q^2)+Q^2}\ .
\ea

Here we want to point out that the two-point functions $\Pi^M_P$ 
and
$\Pi_P$ suffer from the same ambiguity (via its dependence on $g_S$)
as the quark-antiquark one point-function (see discussion at the end
of section \ref{second}) when compared with the $\chi$PT results. 

\subsection{The scalar mixed sector}
\rlabel{scalar0}
Here we have to extend the analysis of \rcite{BRZ} to include possible
mixing effects.
This can be done in the same way as in the previous subsection with the
result (with flavour subscripts $ij$ suppressed),
\ba
\rlabel{62}
{\Pi}_V^{(0)}(Q^2)&=& {1\over \Delta_S (Q^2)} \left[
(1-g_S
\overline{\Pi}_S(Q^2))\overline{\Pi}_V^{(0)}(Q^2) 
+ g_S{(\overline{\Pi}_S^M(Q^2))}^2
\right] \ ,
\\ \rlabel{63}
{\Pi}_S^M(Q^2)&=&{1\over \Delta_S (Q^2)}  {\overline\Pi}_S^M(Q^2)\ ,
\\ \rlabel{64}
 {\Pi}_S(Q^2)&=& {1\over \Delta_S (Q^2)} \left[
(1-g_V\overline{\Pi}_V^{(0)}(Q^2))\overline{\Pi}_S (Q^2) +
 g_V{(\overline{\Pi}^M_S(Q^2))}^2
\right] \ , \\ \rlabel{65}
\Delta_S (Q^2)&=&\left( 1-g_V\overline{\Pi}_V^{(0)}(Q^2)\right)
 \left( 1 -g_S\overline{\Pi}_S(Q^2)\right)   - g_S g_V{\left( 
\overline{\Pi}^M_S(Q^2)\right) }^2 .
\ea
To rewrite this in a simple fashion we would again like to expand
$\Delta_S$ in a simple pole like fashion. Using the identities 
for the one-loop
two-point functions this can almost be done, we obtain
\ba
\rlabel{deltas}
\Delta_S(Q^2)_{ij}&=&\frac{g_S\ovpi_P^M(Q^2)_{ij}}{M_i+M_j}
\left(\left(M_i+M_j\right)^2 + g_A(Q^2)_{ij} m_{ij}^2(Q^2)
+ Q^2 \right)
\nonumber\\\hspace*{0.5cm} &+& 
\ovpi^{(0)}_V(Q^2)_{ij}\left( Q^2 
g_S - g_V \frac{m_i-m_j}{M_i-M_j}\right)\ .
\ea
It can be seen that in the diagonal case a simple 
expression for the scalar meson pole can be found,
\be
\rlabel{ms}
\left.{M_S^2(-M_S^2)}\right|_{m_i=m_j} = 
(M_i+M_j)^2 + g_A(-M_S^2)_{ii}  m_{ii}^2(-M_S^2)\ .
\ee
The expression for the scalar two-point function $\Pi_S (Q^2)$
is in this  case 
\ba \rlabel{pisca}
\left.{\Pi_S(Q^2)}\right|_{m_i=m_j} 
&=& \left\{
-\frac{1}{g_S} + \frac{g_A(Q^2)_{ij} \left(M_i+M_j\right)^2}
{2 f_{ij}^2(Q^2)} \frac{1}{g_S^2}\frac{1}{M_S^2(Q^2)+Q^2}\right\}
_{m_i=m_j} \ .
\ea
So in the
diagonal case a simple relation between the scalar mass, the constituent 
masses and the pseudoscalar mass remains valid to all orders in the masses.
In this case $\Pi_V^{(0)}=\Pi^M_S=0$. 

For the off-diagonal case, i.e. $m_i \ne m_j$, 
 the corresponding expressions
for $\Pi_V^{(0)}$, $\Pi_S^M$ and $\Pi_S$ can be obtained from eqs.
\rref{62}-\rref{65} and the explicit $\ovpi$ functions in appendix 
\ref{AppC}. 
There is a small shift in the pole compared to eq. \rref{ms} for the case 
$m_i\ne m_j$. From appendix \ref{AppC}, in 
eq. \rref{piv0}, it can be seen that 
$\ovpi^{(0)}_V$ itself has a zero close to a value
 of $Q^2 = M_S^2$ of eq. \rref{ms}.  In addition
$\ovpi^{(0)}_V$ is suppressed by $\left(M_i-M_j\right)^2/Q^2$. 
Therefore the value of the pole in the off-diagonal case is not too
far from that in eq. \rref{ms}.

Here we want to point out that (as in the mixed pseudoscalar sector)
the two-point functions $\Pi_V^{(0)}$, $\Pi^M_S$ and
$\Pi_S$ suffer from the same ambiguity (via its dependence on $g_S$)
as the quark-antiquark one point-function (see discussion at the end
of section \ref{second}) when compared with the $\chi$PT results. 

\subsection{Weinberg Sum Rules}
\rlabel{WSRsec}
The Weinberg Sum Rules are general restrictions on the short-distance behaviour
of various two-point functions \rcite{WeinSR}. 
They were first discussed within QCD in ref.
\rcite{Floratos}. A low-energy model of QCD should have a behaviour at
intermediate energies that matches on reasonably well with the QCD behaviour.
The general behaviour should be ($\Pi_{LR} \equiv \Pi_V - \Pi_A$.)
\ba
\rlabel{wsr1}
\lim_{Q^2\to\infty}\left(Q^2 \Pi^{(0+1)}_{LR}(Q^2)\right) = 0
&&\mathrm{First WSR} \,,
\\ \rlabel{wsr2}
\lim_{Q^2\to\infty}\left(Q^4 \Pi^{(1)}_{LR}(Q^2)\right) = 0&&
\mathrm{Second WSR} \, ,
\\ \rlabel{wsr3}
\lim_{Q^2\to\infty}\left(Q^4 \Pi^{(0)}_{LR}(Q^2)\right) = 0&&
\mathrm{Third WSR}\ .
\ea
Let us review first the QCD behaviour of these Sum Rules.
In  the large $N_c$ limit the three WSRs are 
theorems of QCD in the chiral limit (i.e., ${\cal M} \to 0$).
The first WSR is still fulfilled in the large $N_c$ limit with
non-vanishing current quark masses. However the second and the
third ones are violated as follows \rcite{Pascual},
\ba
\lim_{Q^2\to\infty}\left(Q^4 \Pi^{(1)}_{LR}(Q^2)\right) &=&
- \lim_{Q^2\to\infty}\left(Q^4 \Pi^{(0)}_{LR}(Q^2)\right) 
\nonumber \\ &=&
 2 \left( m_i\langle \bar q_j q_j  \rangle +
 m_j \langle \bar q_i q_i \rangle \right) \ .
\ea

As shown in \rcite{BRZ} the class of ENJL-like models does satisfy the
three WSRs in the chiral limit. 
We shall now check how well this does in the case
of explicit breaking of chiral symmetry.

The high-energy behaviour  of the two-point functions
$\Pi^{(0,1)}_{V,A}$ needed for the three WSRs can be easily 
obtained from the
expressions in sections \tref{vector1},
\tref{axialvector1}, \tref{pseudo0} and \tref{scalar0}. 
The first and second WSRs are satisfied in these ENJL-like models
even with non-vanishing and all different current quark-masses.
The  high energy behaviour ($Q^4$) 
of these models is thus too strongly
suppressed for $\Pi^{(1)}_{LR}(Q^2)$ to reproduce the QCD behaviour
in the second WSR.
The third one is violated as in QCD and one has
\be
\lim_{Q^2\to\infty}\left( Q^4\Pi_{LR}^{(0)}(Q^2)\right)
 = \frac{\dis 2}{\dis g_S} \left(m_i M_j + m_j M_i\right) \ .
\ee

Let us now see what relations between low-energy hadronic
couplings do these Sum Rules imply for this ENJL cut-off model. 
In the equal mass sector, $m_i = m_j \ne 0 $, one has  
\ba
\rlabel{wsrel}
f_V^2 M_V^2 = f_A^2 M_A^2 + f^2_\pi \, , \\
f_V^2 M_V^4 = f_A^2 M_A^4 \, .
\ea
Remember that in QCD one has in this case
\ba
\rlabel{wsqcd}
f_V^2 M_V^2 = f_A^2 M_A^2 + f^2_\pi \, , \\
f_V^2 M_V^4 = f_A^2 M_A^4 + m^2_\pi f^2_\pi\, .
\ea
In the off-diagonal case, $m_i\ne m_j$, the situation becomes 
a lot more complicated.  However, since the off-diagonal
part is suppressed by $(M_i-M_j)^2/Q^2$ one does not expect
qualitatively different results.

\subsection{Some numerical results}
\rlabel{numbers}
As can be seen from the explicit formulas the change with respect to
ref. \rcite{BRZ} is in most cases a (small) 
shift in the two-point function mass pole positions.
Therefore we do not plot too many of the two-point functions. 
As numerical input we use for $G_S$, $G_V$ and $\Lambda_\chi$ the values from
fit 1 in ref. \rcite{BBR}. These are $\Lambda_\chi = 1.160$ GeV and
$G_S = 1.216$. The value of $g_A(Q^2=0)$ there was 0.61. This is
$G_V = 1.263$. For the current quark masses we use the value of the quark
mass for ${\overline m}\equiv m_u=m_d$ that reproduces the physical 
neutral pion and kaon masses. With
the other parameters as fixed above this is ${\overline m}=3.2$ MeV
and $m_s/{\overline m}=26$.  

As an example
we have plotted the inverse of 
the transverse vector two-point function in eq. \rref{transvec} in 
figure \ref{FigV1} for the values of $G_S$ and $\Lambda_\chi$ 
corresponding above mentioned.
\begin{figure}
\rotate[r]{\epsfysize=13.5cm\epsfxsize=8cm\epsfbox{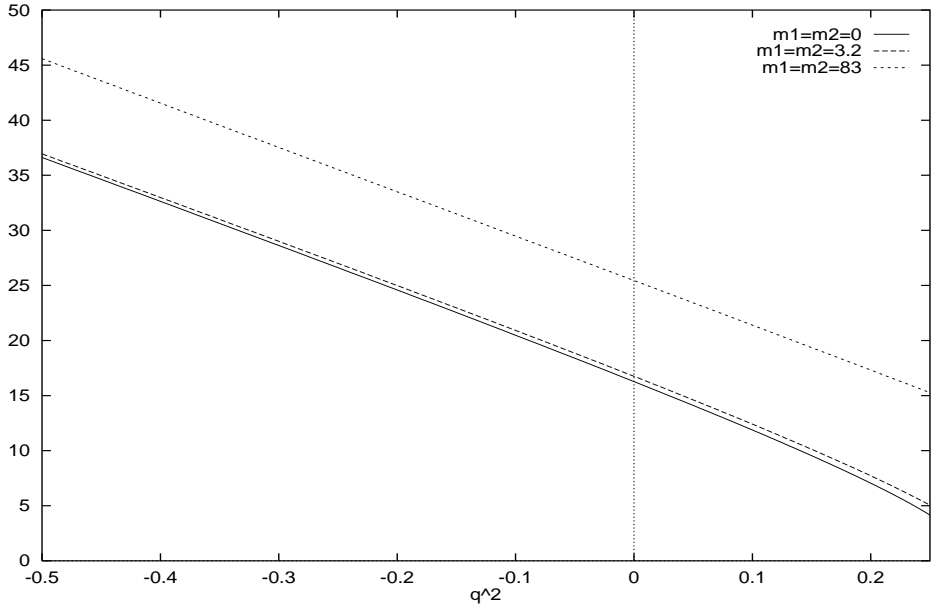}}
\caption{The inverse of the transverse vector 
two-point function for equal
quark masses in the chiral limit, i.e. ${\cal M} \to 0$; 
 for the $\rho$ meson, i.e. $m_1=m_2=3.2$ MeV
and for the $\phi$ meson, i.e. $m_1=m_2= 83$ MeV.
The units of $q^2$ are GeV$^2$}
\rlabel{FigV1}
\end{figure} 
The full curve is the result in the chiral
limit (${\cal M} \to 0$) and the 
dashed is the result with $m_i=m_j=\overline m$ the value above. 
The reason we have 
plotted the inverse will become clear in section \tref{VMD}.
We also show the inverse for $m_i=m_j=m_s$ the value above
in the short-dashed curve.
To show the result for unequal quark masses we have plotted in 
figure \tref{FigV2} the transverse vector
two-point function itself for the chiral
limit  case and for the $\bar u s$ case with $m_s$ and 
$\overline m$ above. Notice that the two-point function now has 
a kinematical pole at $q^2=0$. 
\begin{figure}
\rotate[r]{\epsfysize=13.5cm\epsfxsize=8cm\epsfbox{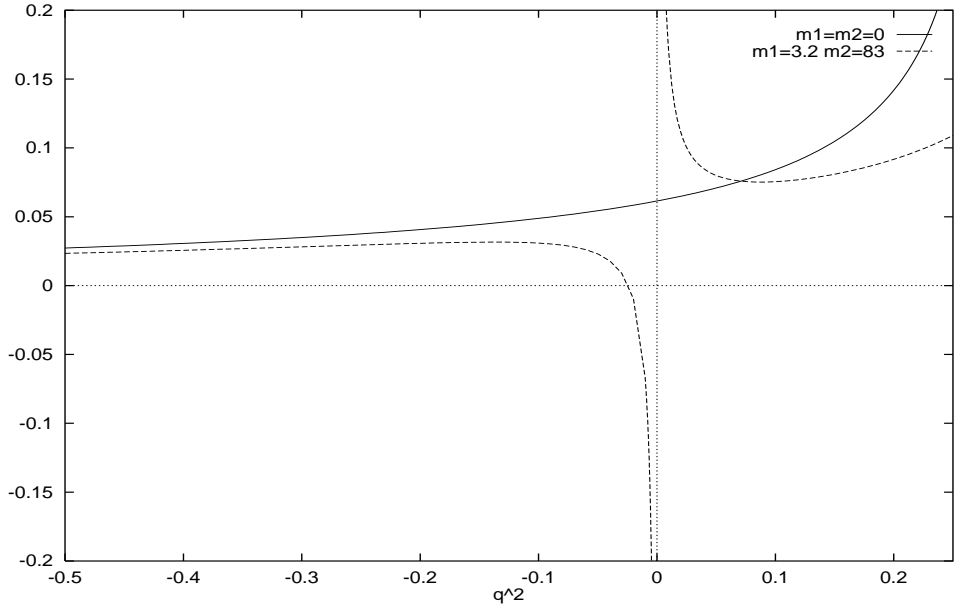}}
\caption{The transverse vector-two-point function for the chiral limit
and for unequal quark masses, $m_1 = \overline m$ and $m_2 = m_s$. 
Note the kinematical pole at $q^2 = 0$. The units of $q^2$ are GeV$^2$.}
\rlabel{FigV2}
\end{figure}

We have also plotted in figure
 \ref{Figmpi} for the parameters quoted above the
dependence of the pion mass on $Q^2$. 
Since $f_{ij}^2 m_{ij}^2$ is a constant,
see eq. \rref{mpi} this is also the $Q^2$ dependence of the
inverse of the $f_{ij}$ decay constant squared.
\begin{figure}
\rotate[r]{\epsfysize=13.5cm\epsfxsize=8cm\epsfbox{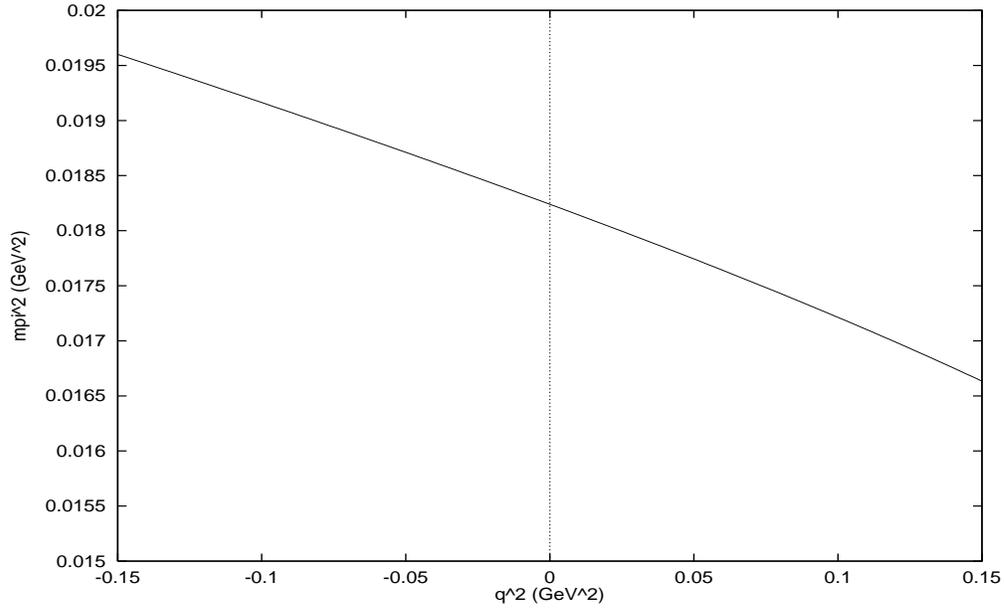}}
\caption{The running pseudoscalar mass squared, $m_{ij}^2(-q^2)$, as a
function of $q^2$ for $m_i = m_j = 3.2$ MeV.}
\rlabel{Figmpi}
\end{figure}

Let us make some comments on the corrections we find to the GMOR
relation \rref{mpi} in this model.
The corrections to the GMOR relation \rcite{Gell} can be
calculated here in an analogous expansion to the one in 
$\chi$PT.
Then the GMOR relation can be written as follows 
\rcite{GL} (for the diagonal flavour case, i.e. $i=j$)
\ba
2 m_i \langle {\overline q}_i q_i \rangle &=& - m_{ii}^2(-q^2) 
f_{ii}^2(-q^2) \left(1 - 4 \frac{\dis m_{ii}^2(0)}
{\dis f_{ii}^2(0)} \left( 2 L_8 - H_2 \right) 
+ {\cal O} (p^6)\, \right) \ . \nonumber \\
\ea
Here we have included all the chiral
corrections to the quark condensate, to the pion mass
and to the pion decay constant in their respective values.
Then the remaining is a correction to the GMOR relation.
We have also calculated this correction in this model
and it turns out to be 
\ba
4 \frac{\dis m_{ii}^2(0)}
{\dis f_{ii}^2(0)} \left( 2 L_8 - H_2 \right)
+ {\cal O} (p^6)
&=& \frac{\dis m_i}{\dis M_i} \, .
\ea 
Notice that the r.h.s. contains all the orders
in the $\chi$PT expansion in the large $N_c$ limit.
Numerically, this correction is around 1 $\%$
for pions and 20 $\%$ for kaons and  approximately 
agrees with the one found in QCD Sum Rules \rcite{Narison}.
For the combination of ${\cal O} (p^4)$ couplings
$2 L_8 - H_2$ in this model we get 
\ba \rlabel{h2}
2 L_8 - H_2 &=& \frac{\dis N_c}{\dis 16 \pi^2}
\, \frac{\dis g_A^2(0)}{\dis 2} \frac{\dis
\Gamma^2(0,\epsilon)}{\dis \Gamma(-1,\epsilon)} \, ,
\ea
with $\epsilon=M^2/\Lambda_\chi^2$ and $M$ the constituent
quark mass in the chiral limit. (For definitions of the incomplete
Gamma functions $\Gamma(n,\epsilon)$ see appendix \tref{GAMMA}.)
 The expression in \rref{h2} is the  one consistent with
the use of Ward identities to sum the infinite string
of constituent quark bubbles. Numerically we get
$2 L_8 - H_2 \simeq
1.3 \cdot 10^{-3}$ for the input parameters above.
  This differs from the one found
at the one-loop level, in this same model in ref. 
\rcite{BBR} (see footnote $^1$), numerically they find
$2 L_8 - H_2 = 0.2 \cdot 10^{-3}$.

\setcounter{equation}{0}
\section{Some three-point functions}
\rlabel{threep}
\subsection{VPP with the use of the Ward identities}
\rlabel{vppsub}
In this subsection we calculate the 
Vector Pseudoscalar Pseudoscalar (VPP) three-point
function to all orders in $\chi$PT
using the same type of methods as those used for the two-point
functions. The three-point function we calculate is the following
\be
\rlabel{VPP}
\Pi_\mu^{VPP}(p_1,p_2) \equiv
i^2 \int {\rm d}^4x \int {\rm d}^4y e^{i(p_1\cdot x + p_2 \cdot y)}
\langle 0 | T\left(V^{ij}_\mu(0)P^{kl}(x)P^{mn}(y)\right)| 
0 \rangle \ .
\ee
Where $i,j,k,l,m$ and $n$ are flavour indices.
In the limit of large $N_c$ the flavour structure is limited because
of Zweig's rule (this flavour structure is general for any 
three-point function of three quark currents),
\be
\Pi_\mu^{VPP}(p_1,p_2) \equiv
\Pi^{+}_\mu(p_1,p_2)_{ikm}  \delta_{il}\delta_{kn}\delta_{mj}
+\Pi^{-}_\mu(p_1,p_2)_{ikm} \delta_{in}\delta_{kj}\delta_{ml}\ .
\ee
Bose symmetry requires that
\be 
\rlabel{pi+}
\Pi^{+}_\mu(p_1,p_2)_{ikm} = \Pi^{-}_\mu(p_2,p_1)_{imk}\ .
\ee
The three-point function $\Pi^{VPP}_\mu (p_1,p_2)$
 can then be simply calculated by only taking one
particular flavour combination.
Finally we can use Lorentz-invariance to rewrite
\be
\Pi^{+}_\mu(p_1,p_2)_{ikm} = p_{1\mu} \Pi^A_{ikm}(p_1^2,p_2^2,q^2)
+ p_{2\mu} \Pi^B_{ikm}(p_1^2,p_2^2,q^2)\ ,
\ee
where we have defined $q \equiv p_1 + p_2$.

We shall limit ourselves to the vector diagonal case, i.e.
 $m_i = m_j$.
In the vector off-diagonal case there will also
be non-trivial mixings with the 
scalar-pseudoscalar-pseudoscalar three-point function. 
Here a relatively
simple Ward identity for this three-point function can be derived from
$\partial^\mu V^{ij}_\mu = 0$ and the equal-time commutation relations.
It is
\be
\rlabel{WVPP}
q^\mu \Pi^{+}_\mu (p_1,p_2)_{ikm} =
-\Pi_P(-p_1^2)_{ki}+\Pi_P(-p_2^2)_{mk}\ .
\ee
So the Ward identity relates the three-point function to a combination of
two-point functions. This determines one of the two functions
$\Pi^A, \Pi^B$ in terms of the other. The Ward identity gives, 
for instance, the following constraint (for $p_1^2=p_2^2$ and $i=m$)
\be
\rlabel{WAB}
\Pi^B_{iki}(p^2,p^2,q^2)= -\Pi^A_{iki}(p^2,p^2,q^2)\ .
\ee

 The type of graphs that need to be summed
are depicted in figure \ref{Fig3graf}.
\begin{figure}
\begin{center}
%
%
%
\thicklines
\setlength{\unitlength}{1mm}
\begin{picture}(140.00,80.00)(10.,0.)
\put(133.00,18.50){\oval(15.00,10.00)}
\put(136.25,51.75){$\bigotimes$}
\put(118.00,18.50){\oval(15.00,10.00)}
\put(95.50,18.50){\circle*{2.00}}
\put(103.00,18.50){\oval(15.00,10.00)}
\put(136.00,13.50){\vector(-1,0){3.50}}
\put(121.50,13.50){\vector(-1,0){5.00}}
\put(105.00,13.50){\vector(-1,0){3.00}}
\put(101.50,23.50){\vector(1,0){3.00}}
\put(116.00,23.50){\vector(1,0){3.50}}
\put(131.50,23.50){\vector(1,0){2.00}}
\put(32.50,35.00){\oval(15.00,10.00)}
\put(47.50,35.00){\oval(15.00,10.00)}
\put(23.00,33.50){$\bigotimes$}
\put(62.50,35.00){\oval(15.00,10.00)}
\put(70.00,35.00){\circle*{2.00}}
\put(29.50,40.00){\vector(1,0){3.50}}
\put(44.00,40.00){\vector(1,0){5.00}}
\put(60.50,40.00){\vector(1,0){3.00}}
\put(64.00,30.00){\vector(-1,0){3.00}}
\put(131.00,53.00){\oval(14.00,10.00)}
\put(123.50,53.00){\circle*{2.00}}
\put(116.50,53.00){\oval(14.00,10.00)}
\put(109.50,53.00){\circle*{2.00}}
\put(102.50,53.00){\oval(14.00,10.00)}
\put(133.50,48.00){\vector(-1,0){3.50}}
\put(120.00,48.00){\vector(-1,0){4.50}}
\put(104.50,48.00){\vector(-1,0){3.00}}
\put(101.00,58.00){\vector(1,0){3.00}}
\put(115.00,58.00){\vector(1,0){3.50}}
\put(129.50,58.00){\vector(1,0){2.00}}
\put(110.50,18.50){\circle*{2.00}}
\put(96.00,18.50){\circle*{0.00}}
\put(95.50,53.00){\circle*{2.00}}
\put(82.50,36.00){\oval(25.00,55.00)}
\put(35.50,30.00){\vector(-1,0){3.50}}
\put(95.00,35.50){\vector(0,-1){3.50}}
\put(82.50,63.50){\vector(1,0){3.50}}
\put(83.50,8.50){\vector(-1,0){3.50}}
\put(49.50,30.00){\vector(-1,0){3.50}}
\put(125.50,18.50){\circle*{2.00}}
\put(138.25,16.75){$\bigotimes$}
\put(10.75,32.75){$\gamma_\mu$}
\put(72.00,33.00){$\gamma_\nu$}
\put(78.00,52.75){$\gamma_\alpha\gamma_5,i\gamma_5$}
\put(78.75,16.50){$\gamma_\beta\gamma_5,i\gamma_5$}
\put(139.00,48.75){$i\gamma_5$}
\put(141.00,11.00){$i\gamma_5$}
\put(141.00,43.00){\vector(1,0){5.75}}
\put(142.25,5.00){\vector(1,0){6.25}}
\put(139.75,44.75){$p_1$}
\put(141.00,7.00){$p_2$}
\put(107.00,69.75){Tail II}
\put(109.25,3.50){Tail III}
\put(32.25,49.75){Tail I}
\put(44.25,44.25){i}
\put(43.75,23.75){m}
\put(81.25,67.75){i}
\put(115.00,62.00){i}
\put(115.50,41.25){k}
\put(99.25,34.75){k}
\put(117.00,27.75){k}
\put(117.25,8.50){m}
\put(81.75,3.25){m}
\put(15.50,25.50){\vector(1,0){7.25}}
\put(15.75,28.50){$q$}
\put(55.00,35.00){\circle*{2.00}}
\put(40.50,35.00){\circle*{2.00}}
\end{picture}
\caption{The graphs that need to be summed in the large $N_c$ limit for 
the Vector-Pseudoscalar-Pseudoscalar three-point function. See text for
explanation.}
\rlabel{Fig3graf}
\end{center}
\end{figure}
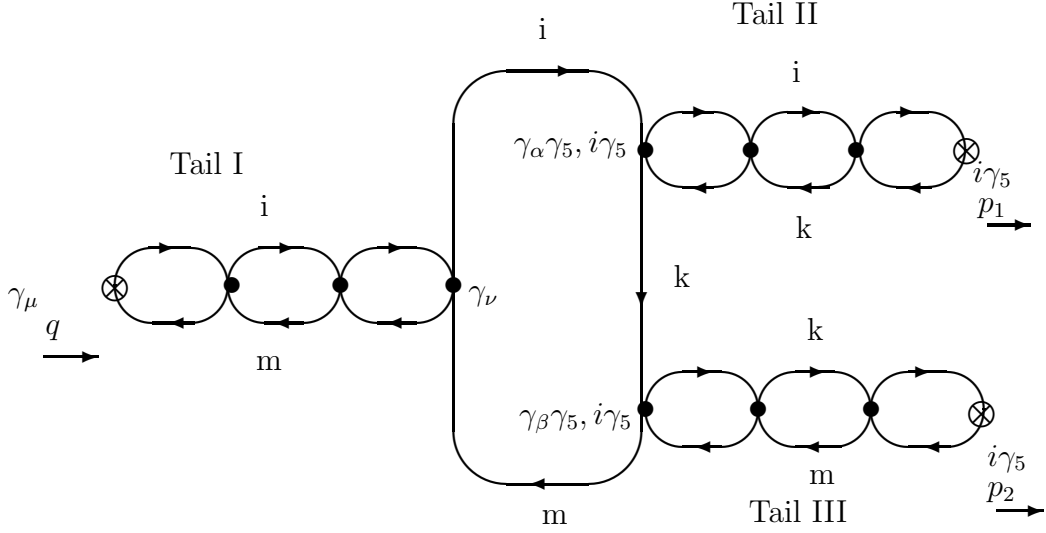 
Each of the three tails  here 
is  the diagram in figure  \tref{Fig2pt}a with the same explanation
as there.  
We have there depicted one particular flavour combination.
This is the one that corresponds to the 
function  $\Pi^{+}_\mu$ given above. 
The $i,k,m$ written above the lines are the flavours of each line.

All graphs are formed by having the tails summed over 0, 1, 2, 
$\cdots$, $\infty$
loops connected by four-fermion couplings. These then couple to 
the one-loop
three-point function (or vertex) $\ovpi^{+}_\mu$,
with various possibilities for the insertion in the
three-point vertex.
These possibilities for the $\gamma$-matrices are written
in figure \ref{Fig3graf} inside the main loop.
 
In this figure the left-hand side depicts the insertion of
 the current $V^{ij}_\mu (0)$ and 
Tail I is the connection to this current. On the end connecting to
the one-loop three-point function it is only nonzero for another
vector insertion since in the diagonal case we consider, the mixed 
vector--scalar two-point function vanishes.
It expression is given by
\be 
g_{\mu\nu} + \frac{-8\pi^2 G_V}{N_c\Lambda^2_\chi}
\Pi^V_{\mu\nu}(-q^2)_{mi}\ .
\ee
Here the first term comes from where the external current directly connects
to the one-loop three-point function and the second term is with the
two-point function in between. The sum of both is
\be
\frac{g_{\mu\nu}M_V^2(-q^2)_{mi}-q_\mu q_\nu}
{M_V^2(-q^2)_{mi}-q^2}.
\ee

A similar discussion can be done for Tail II and Tail III. 
First we have the 
insertion of the current $P^{kl}(x)$ 
at the external end. On the end connecting
to the one-loop three-point function we can have $i\gamma_5$ or an 
axial-vector insertion since the mixed axial-vector--pseudoscalar 
two-point function is nonzero. The $i\gamma_5$ insertion tail is :
\ba
\rlabel{ga5}
1 + \frac{4\pi^2 G_S}{N_c\Lambda^2_\chi}\Pi_P(-p_1^2)_{ki}
\nonumber \\ \hspace*{1cm}
= \frac{\left(M_k+M_i\right)^2}{2 g_S f_{ki}^2(-p_1^2) 
\left(m_{ki}^2(-p_1^2) - p_1^2\right)}\ .
\ea
For the connection with the axial-vector insertion it is instead
\ba
\rlabel{axiv}
\frac{8\pi^2 G_V}{N_c\Lambda^2_\chi}
i p_1^\alpha \Pi^M_P(-p_1^2)_{ki} \nonumber 
\\ \hspace*{1cm} =
\frac{ip_1^\alpha}{2 f_V^2 M_V^2}
\frac{\left(M_k+M_i\right)}{g_S 
\left(m_{ki}^2(-p_1^2) - p_1^2\right)}\ .
\ea
The combination $f_V^2 \, M_V^2$ is here flavour and $p_1^2$ independent,
it is the combination in eq. \rref{fvmv} with 
$\ovpi^{(0)}_V(Q^2)_{ij}=0$ since we are in the diagonal flavour case. 
The way both these types of insertions
can appear due to the tail are how within this formulation the mixing
of pseudoscalar and axial-vector degrees comes about. These will 
be described by factors of $g_A^2$ (see below).
Tail III is identical to Tail II with the substitutions $p_1\to p_2$ and
$i,k\ \to\ k,m$.

The full expression for $\Pi^+_\mu$ is
\be
\renewcommand{\arraystretch}{1.5}
\begin{array}{l}
\rlabel{fvpp} \dis
\Pi^{+\mu}(p_1,p_2) = \left\{
 g^{\mu\nu} + \frac{-8\pi^2 G_V}{N_c\Lambda^2_\chi}
\Pi^{V\mu\nu}(-q^2)_{mi} \right\} \nonumber \\ \dis
\hspace*{1cm}
\times \left\{ \ovpi^+_\nu(p_1,p_2) \left( 
1 + \frac{4\pi^2 G_S}{N_c\Lambda^2_\chi} \Pi_P(-p_1^2)_{ki} \right)
\left( 1 + \frac{4\pi^2 G_S}{N_c\Lambda^2_\chi} \Pi_P(-p_2^2)_{mk}
\right) \right. \nonumber \\ \dis \hspace*{1cm} 
+ \ovpi^{VPA}_{\nu\beta}(p_1,p_2) 
\left(1 + \frac{4\pi^2 G_S}{N_c\Lambda^2_\chi} \Pi_P(-p_1^2)_{ki} 
\right) \left( \frac{8\pi^2 G_V}{N_c\Lambda^2_\chi}
i p_2^\beta \Pi^M_P(-p_2^2)_{mk} \right) \nonumber \\ \dis
\hspace*{1cm}
+  \ovpi^{VAP}_{ \nu\alpha}(p_1,p_2)
\left(\frac{8\pi^2 G_V}{N_c\Lambda^2_\chi}
i p_1^\alpha  \Pi^M_P(-p_1^2)_{ki} \right) 
\left(1 + \frac{4\pi^2 G_S}{N_c\Lambda^2_\chi}
\Pi_P(-p_2^2)_{mk} \right)  \nonumber \\ \dis
\hspace*{1cm} + \left. \ovpi^{VAA}_{\nu\alpha\beta}(p_1,p_2)
\left( \frac{8\pi^2 G_V}{N_c\Lambda^2_\chi}
i p_1^\alpha \Pi^M_P(-p_1^2)_{ki} \right)
\left( \frac{8\pi^2 G_V}{N_c\Lambda^2_\chi}
i p_2^\beta \Pi^M_P(-p_2^2)_{mk} \right) \right\} \, .
\nonumber \\
\end{array}
\renewcommand{\arraystretch}{1.5}
\ee
Where the one-loop three-point functions $\ovpi^{VPA}_{\mu\nu}$,
$\ovpi^{VAP}_{\mu\nu}$ and $\ovpi^{VAA}_{\mu\nu\alpha}$
are the one fermion-loop result for
\ba
\rlabel{VPA}
\Pi_{\mu\nu}^{VPA}(p_1,p_2) &\equiv &
i^2 \int {\rm d}^4x \int {\rm d}^4y e^{i(p_1\cdot x + p_2 \cdot y)}
\langle 0 | T\left(V^{im}_\mu(0)P^{ki}(x)A^{mk}_\nu (y)\right)| 
0 \rangle \, , \nonumber \\ \\
\rlabel{VAP}
\Pi_{\mu\nu}^{VAP}(p_1,p_2) &\equiv &
i^2 \int {\rm d}^4x \int {\rm d}^4y e^{i(p_1\cdot x + p_2 \cdot y)}
\langle 0 | T\left(V^{im}_\mu(0)A^{ki}_\nu(x)P^{mk}(y)\right)| 
0 \rangle \, , \nonumber \\ \\
\rlabel{VAA}
\Pi_{\mu\nu\alpha}^{VAA}(p_1,p_2) &\equiv &
i^2 \int {\rm d}^4x \int {\rm d}^4y e^{i(p_1\cdot x + p_2 \cdot y)}
\langle 0 | T\left(V^{im}_\mu(0)A^{ki}_\nu(x)A^{mk}_\alpha
 (y)\right)| 0 \rangle \, . \nonumber \\
\ea

To obtain the full expression in eq. \rref{fvpp}
it now remains to calculate these 
VPP, VAP, VPA and VAA one-loop three-point functions (or vertices).
 The axial-vector
ones always come multiplied with the relevant momentum. So we
always have the scalar products 
$p_1\cdot A^{ki}(x)$ and $p_2\cdot A^{mk} (y)$.
 That means that using the
Ward identities we can relate the VAA, VAP, VPA to 
the VPP one plus possibly two-point function
terms resulting from equal time commutators.

These Ward identities 
are (remember we assume $M_i = M_j$ here).
\ba
i p_1^\nu \ovpi^{VAA}_{\mu \nu \alpha} (p_1,p_2) &=&
- \left(M_k + M_i\right) \ovpi^{VPA}_{\mu \alpha}(p_1,p_2)
\nonumber\\
&& - \ovpi^V_{\mu\alpha}(-q)_{mi} +
 \ovpi^A_{\mu\alpha}(-p_2)_{mk}\ ;
\\
i p_1^\nu \ovpi^{VAP}_{\mu \nu} (p_1,p_2) &=&
- \left(M_k + M_i\right) \ovpi^+_\mu (p_1,p_2) \nonumber\\
&&  + i \, \ovpi_{P \mu}(-p_2)_{mk}\ .
\ea
The other needed ones can be derived from this using Bose-symmetry.
Notice that there is no contribution here from the flavour chiral
anomaly (see eq. \rref{anomW}). 

We can now use these identities to obtain the final result for the three-point
function we want. The terms which after the use of the one-loop identities
above are proportional to VPP can be combined into a simple form
using $g_A(p^2)$. The result is (we have $M_i = M_j$ and $j=m$
in this flavour configuration).
\be 
\renewcommand{\arraystretch}{1.5}
\begin{array}{l} \dis
\rlabel{VPPres} 
\Pi^{+ \mu}(p_1,p_2)\,=\nonumber \\ \dis
\hspace*{3cm} \left(\frac{\left(M_i+M_k\right)^4}
{4 g_S^2 f_{ki}^2(-p_1^2)f_{mk}^2(-p_2^2)} \right) 
\left(\frac{g^{\mu\nu} M_V^2(-q^2)_{mi} - q^\mu q^\nu}
{M_V^2(-q^2)_{mi} - q^2}\right) \nonumber \\ \dis
\times \, \frac{1}{\left( m_{ki}^2(-p_1^2)-p_1^2 \right) 
\left(m_{mk}^2(-p_2^2)-p_2^2 \right)} 
\Bigg\{g_A(-p_1^2)_{ki} g_A(-p_2^2)_{mk} \ovpi^+_\nu (p_1,p_2)
 \nonumber \\ \dis
\hspace*{1cm} + \frac{\left(1-g_A(-p_1^2)_{ki}\right)
\left(1-g_A(-p_2^2)_{mk}\right)}{\left(M_i+M_k\right)^2} \left\{
\left(p_2\cdot q \right) p_{1\nu} - \left( p_1\cdot q \right)
p_{2\nu}\right\} \ovpi^{(1)}_V(-q^2)_{mi}
\nonumber\\ \dis \hspace*{1cm} 
- \frac{g_A(-p_1^2)_{ki}\left(1-g_A(-p_2^2)_{mk}
\right)}{M_i+M_k}p_{1\nu} \ovpi_P^M(-p_1^2)_{ki} \nonumber 
\\ \dis \hspace*{1cm}
+ \frac{g_A(-p_2^2)_{mk}\left(1-g_A(-p_1^2)_{ki}
\right)}{M_i+M_k}p_{2\nu}
\ovpi_P^M(-p_2^2)_{mk} \Bigg\}\ . \nonumber \\
\end{array}
\renewcommand{\arraystretch}{1}
\ee
This result satisfies the Ward identity \rref{WVPP} 
if the one-loop function $\ovpi^+_\mu$ one
satisfies the same one with the one-loop functions. This provides
a rather non-trivial check on the result \rref{VPPres}.

It now only remains to calculate the one-loop form factor
$\ovpi^+_\mu(p_1,p_2)$. We give its expression in 
appendix \ref{AppD}. At this point we can see in  eq. \rref{VPPres}
how far regularization ambiguities affect the result. We first have to 
define the two-point functions. Here all ambiguities are restricted to two
bare functions (see section \tref{twop} for details). 
This three-point function adds one more in general, the 
three-propagator function $I_3(p_1^2,p_2^2,q^2)$ (see explicit expression
in appendix \ref{AppD}). Of course, this one-loop form factor
$\ovpi^+_\mu(p_1,p_2)$, satisfies all the identities eqs. \rref{VPP}
to \rref{WAB} as well. We refer to section \tref{threevpp}
for the definition of the physical vector form factor
after reducing this $VPP$ three-point function. We shall also discuss
there  the VMD limit in this form factor and give some numerics.

The same three-point function can be calculated in Chiral Perturbation
Theory. The result is
\be
\Pi^{+\mu}(p_1,p_2) =
\frac{2 B_0^2 f_{mi}^2}{(m_{ki}^2-p_1^2)(m_{mk}^2-p_2^2)}
\left(p_2-p_1\right)^\mu
\left(1+\frac{2 L_9}{f_{mi}^2}q^2 + {\cal O}(p^6) \right)\ .
\ee
 Pulling out the pion poles (see section \tref{threevpp}
for technical details) and taking the low-energy limit
 and the value of $L_9$ in this class of models our
full result in eq. \rref{VPPres} 
reduces to this, providing one more non-trivial check.
 
\subsection{PVV with a discussion about its Ward identity}
\rlabel{pi0gg}
In this subsection we calculate the Pseudoscalar Vector Vector 
(PVV) three-point function to all orders in $\chi$PT
with the same method as the one used before.
\be
\rlabel{PVV}
\Pi_{\mu\nu}^{PVV}(p_1,p_2) \equiv
i^2 \int {\rm d}^4x \int {\rm d}^4y e^{i(p_1\cdot x + p_2 \cdot y)}
\langle 0 | T\left(P^{ij}(0)V^{kl}_\mu(x)V^{mn}_\nu(y)\right)| 
0 \rangle \ .
\ee
Where $i,j,k,l,m$ and $n$ are flavour indices.
A similar discussion about the structure due to Zweig's rule can be given
as was done before. We do 
the analogous decomposition into $\Pi^+_{\mu\nu}$ and 
$\Pi^-_{\mu\nu}$ 
functions as was done for the three-point function $VPP$
(see previous section).
We shall here restrict ourselves to the case where all
current masses or constituent masses are equal. 
Our main aim in this subsection
is to show how the flavour anomaly \rcite{anomaly} 
affects the use of the  one-loop identities.

The class of graphs
needed here is shown in figure
 \ref{FigPVV} for the $\Pi^+$ flavour combination.
\begin{figure}
\begin{center}
%
%
%
\thicklines
\setlength{\unitlength}{1mm}
\begin{picture}(140.00,80.00)(10.,0.)
\put(133.00,18.50){\oval(15.00,10.00)}
\put(136.25,51.75){$\bigotimes$}
\put(118.00,18.50){\oval(15.00,10.00)}
\put(95.50,18.50){\circle*{2.00}}
\put(103.00,18.50){\oval(15.00,10.00)}
\put(136.00,13.50){\vector(-1,0){3.50}}
\put(121.50,13.50){\vector(-1,0){5.00}}
\put(105.00,13.50){\vector(-1,0){3.00}}
\put(101.50,23.50){\vector(1,0){3.00}}
\put(116.00,23.50){\vector(1,0){3.50}}
\put(131.50,23.50){\vector(1,0){2.00}}
\put(32.50,35.00){\oval(15.00,10.00)}
\put(47.50,35.00){\oval(15.00,10.00)}
\put(23.00,33.50){$\bigotimes$}
\put(62.50,35.00){\oval(15.00,10.00)}
\put(70.00,35.00){\circle*{2.00}}
\put(29.50,40.00){\vector(1,0){3.50}}
\put(44.00,40.00){\vector(1,0){5.00}}
\put(60.50,40.00){\vector(1,0){3.00}}
\put(64.00,30.00){\vector(-1,0){3.00}}
\put(131.00,53.00){\oval(14.00,10.00)}
\put(123.50,53.00){\circle*{2.00}}
\put(116.50,53.00){\oval(14.00,10.00)}
\put(109.50,53.00){\circle*{2.00}}
\put(102.50,53.00){\oval(14.00,10.00)}
\put(133.50,48.00){\vector(-1,0){3.50}}
\put(120.00,48.00){\vector(-1,0){4.50}}
\put(104.50,48.00){\vector(-1,0){3.00}}
\put(101.00,58.00){\vector(1,0){3.00}}
\put(115.00,58.00){\vector(1,0){3.50}}
\put(129.50,58.00){\vector(1,0){2.00}}
\put(110.50,18.50){\circle*{2.00}}
\put(96.00,18.50){\circle*{0.00}}
\put(95.50,53.00){\circle*{2.00}}
\put(82.50,36.00){\oval(25.00,55.00)}
\put(35.50,30.00){\vector(-1,0){3.50}}
\put(95.00,35.50){\vector(0,-1){3.50}}
\put(82.50,63.50){\vector(1,0){3.50}}
\put(83.50,8.50){\vector(-1,0){3.50}}
\put(49.50,30.00){\vector(-1,0){3.50}}
\put(125.50,18.50){\circle*{2.00}}
\put(138.25,16.75){$\bigotimes$}
\put(10.75,32.75){$i\gamma_5$}
\put(72.00,33.00){$i\gamma_5,\gamma_\delta\gamma_5$}
\put(88.00,52.75){$\gamma_\alpha$}
\put(88.75,16.50){$\gamma_\beta$}
\put(139.00,48.75){$\gamma_\mu$}
\put(141.00,11.00){$\gamma_\nu$}
\put(141.00,43.00){\vector(1,0){5.75}}
\put(142.25,5.00){\vector(1,0){6.25}}
\put(139.75,44.75){$p_1$}
\put(141.00,7.00){$p_2$}
\put(107.00,69.75){Tail II}
\put(109.25,3.50){Tail III}
\put(32.25,49.75){Tail I}
\put(44.25,44.25){i}
\put(43.75,23.75){m}
\put(81.25,67.75){i}
\put(115.00,62.00){i}
\put(115.50,41.25){k}
\put(99.25,34.75){k}
\put(117.00,27.75){k}
\put(117.25,8.50){m}
\put(81.75,3.25){m}
\put(15.50,25.50){\vector(1,0){7.25}}
\put(15.75,28.50){$q$}
\put(55.00,35.00){\circle*{2.00}}
\put(40.50,35.00){\circle*{2.00}}
\end{picture}
\caption{The graphs that need to be summed in the large $N_c$
limit for  
the Pseudoscalar-Vector-Vector three-point function. See text
for explanation.}
\rlabel{FigPVV}
\end{center}
\end{figure}
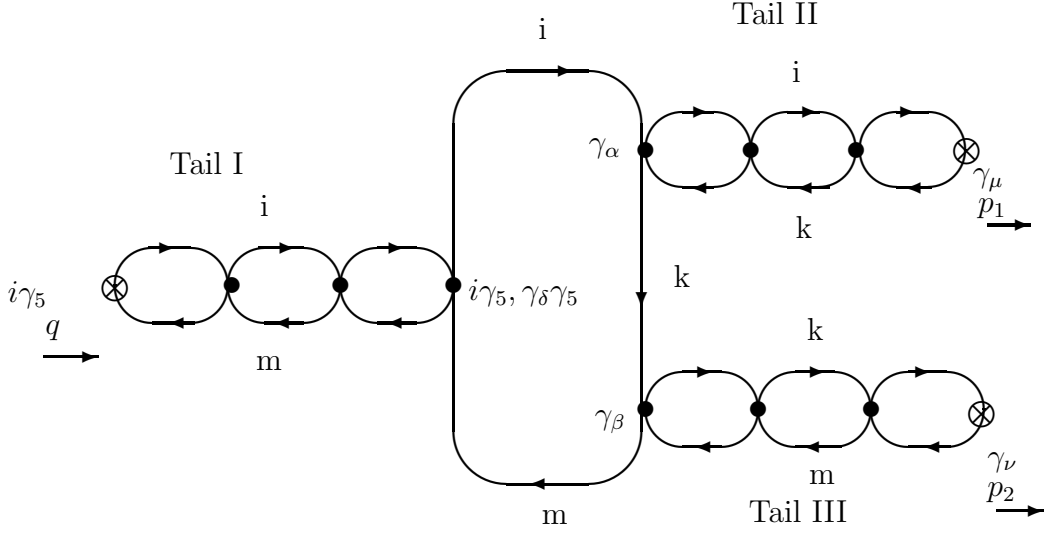
Each of the three tails  here 
is  the diagram in figure  \tref{Fig2pt}a with the same 
explanation as there.  
The vector-like tails, II and III, can be easily summed 
(see discussion for the summation of the vector tail
in the previous section) to obtain the overall
factors
\be
g_{\mu\alpha} + \frac{-8\pi^2 G_V}{N_c\Lambda^2_\chi}
\Pi^V_{\mu\alpha}(-p_1^2)\ 
\ee
and
\be
g_{\nu\beta} + \frac{-8\pi^2 G_V}{N_c\Lambda^2_\chi}
\Pi^V_{\nu\beta}(-p_2^2)\ . 
\ee
Tail I can again couple at the one-loop end to both an axial-vector and
pseudoscalar two-point function. These have the same form as equations
\rref{ga5} and \rref{axiv} in the previous section with $p_1\to q$.  
Summing up the three tails the total result is then

\be
\rlabel{wardan}
\renewcommand{\arraystretch}{1.5}
\begin{array}{l}
\Pi^{+\mu\nu}(p_1,p_2) = \nonumber \\ \dis
 \left( g^{\mu\alpha} + 
\frac{-8\pi^2 G_V}{N_c\Lambda^2_\chi} \Pi^{V \mu\alpha}(-p_1^2)
\right) \left( g^{\nu\beta} + 
\frac{-8\pi^2 G_V}{N_c\Lambda^2_\chi} \Pi^{V \nu\beta}(-p_2^2)
\right) \nonumber \\ \dis
\times \left[ \ovpi^+_{\alpha\beta}(p_1,p_2) \left\{ 1 +
\frac{\dis 4 \pi^2 G_S}{\dis N_c \Lambda_\chi^2}
\Pi_P(-q) \right\}+\ovpi^{AVV}_{\rho \alpha \beta}
(p_1,p_2) \left( \frac{\dis 8 \pi^2 G_V}{\dis N_c
\Lambda_\chi^2} i q^\rho \Pi^M_P (-q)
 \right)\right] \nonumber \\
\end{array}
\renewcommand{\arraystretch}{1}
\ee
with $\ovpi^{AVV}_{\alpha\mu\nu}$ the one-loop result for
the following three-point function
\ba
\rlabel{AVV}
\Pi_{\rho\mu\nu}^{AVV}(p_1,p_2) &\equiv &
i^2 \int {\rm d}^4x \int {\rm d}^4y e^{i(p_1\cdot x + p_2 \cdot y)}
\langle 0 | T\left(A^{im}_\rho(0)V^{ki}_\mu(x)V^{mk}_\nu (y)\right)| 
0 \rangle \, . \nonumber \\
\ea

The main new part here is that at the 
one-loop level we now have to include the anomalous part of 
the Ward identities.
There has been in fact quite some confusion whether this can be done 
consistently. We have shown how this subtraction needs to be done in the
case of ENJL-like models in ref. \rcite{BP}. The anomaly itself in this class
of models is not well defined but a consistent subtraction procedure to obtain
the QCD flavour anomaly can be easily formulated, see ref. \rcite{BP}.
We could in principle use the prescription of ref. \rcite{BP} directly to
obtain the one-loop three-point function.  Here we want to apply
it to the PVV three-point function to all orders in external momenta
and quark masses.
The prescription is essentially
to use the anomalous QCD Ward identities  for the axial current 
consistently. We shall use the scheme where vector currents are 
conserved \rcite{Bardeen}. The subtractions these Ward 
identities impose in order to reproduce 
the correct QCD flavour anomaly
in the ENJL-like models  effective action we are working with, 
lead to the use of the following consistent one-loop
anomalous Ward identity
\ba
\rlabel{anomW}
\partial_\mu A^\mu_{ii}(x) &=& 2 M_i P_{ii}(x)
 + \frac{N_c}{16\pi^2} 
\varepsilon_{\mu\nu\alpha\beta}\left(v^{\mu\nu}_{ki}(x)
v^{\alpha\beta}_{ik}(x) \right. \nonumber
 \\ &+&  {4 \over 3} {\rm d}^\mu a^\nu_{ki}(x)
{\rm d}^\alpha a^\beta_{ik}(x) +  {2 \over 3} i \left\{
v^{\mu\nu}_{ki}(x),a^\alpha_{im}a^\beta_{mk}(x)\right\}
\nonumber \\ &+& \left. 
{8 \over 3} i (a^\alpha v^{\mu\nu}a^\beta)_{ii}(x)
 + { 4 \over 3} (a^\mu a^\nu a^\alpha a^\beta)_{ii}(x)
\right)\, \nonumber \\
{\rm with} \hspace*{2cm}\nonumber \\
 v^{\mu\nu} &\equiv& \partial^\mu v^\nu - \partial^\nu v^\mu
- i \left[v^\mu,v^\nu \right] \, \hspace*{0.5cm}
{\rm and} \nonumber \\
{\rm d}^\mu a^\nu &\equiv& \partial^\mu a^\nu 
- i \left[v^\mu,a^\nu \right] \, . \nonumber \\
\ea
Where $v^\mu$ and $a^\mu$ are the external vector and axial-vector
sources defined in eq. \rref{QCD}.
As shown in ref. \rcite{BP}, when using this Ward identity,
the fact that the anomalous
part in eq. \rref{anomW} only contains external fields
amounts to keeping the usual Wess-Zumino term 
\rcite{WZW} (the only one
of ${\cal O} (p^4)$ in the chiral counting) for 
couplings of pseudoscalar type via $G_S$ to external fields
 but when there
are couplings to spin-1 fields via the $G_V$ term,
only the local chiral invariant part of the full term remains. 
We have checked that the form of the 
action given in \rcite{BP} yields the same result as the 
one given below. 

So when we use the one-loop anomalous 
Ward identity in eq. \rref{anomW}
to reduce the right-hand side of Tail I to
a part with only pseudoscalar couplings to the one-loop
vertex, we obtain a local chiral invariant result plus
an extra part where Tail I  couples directly to the
external vector sources $v^{kl}_\mu(x) v^{mn}_\nu(y)$. 
This extra part is
of order $p^4$ and is the subtraction the anomalous
Ward identity  
imposes to obtain the correct QCD flavour anomaly.

The full result in terms of the one-loop $\ovpi^+_{\mu\nu}$
three-point function is given by
\ba
\rlabel{PVVr1}
\Pi^+_{\mu\nu}(p_1,p_2) &=& \ovpi^+_{\mu\nu}(p_1,p_2)
\left(\frac{\dis  
M_V^2(-p_1^2)_{ii} M_V^2(-p_2^2)_{ii}}{\dis \left(
M_V^2(-p_1^2)_{ii} - p_1^2\right) \left( M_V^2(-p_2^2)_{ii}
- p_2^2 \right)} \right) \nonumber \\ &\times&
\left\{ 1 + \frac{\dis 4 \pi^2 G_S}{\dis N_c \Lambda_\chi^2}
\Pi_P(-q) - \frac{\dis 8 \pi^2 G_V}{\dis N_c \Lambda_\chi^2}
2 M_i \Pi^M_P (-q) \right\} \nonumber \\
&+& \ovpi^+_{\mu\nu}(p_1,p_2)\Bigg|_{p_1^2=p_2^2=q^2=0}
\frac{\dis 8 \pi^2 G_V}{\dis N_c \Lambda_\chi^2}
2 M_i \Pi^M_P (-q)  \, .
\ea
Where the one-constituent quark loop function $\ovpi^+_{\mu\nu}$
is given by 
\ba
\rlabel{onam}
\ovpi^+_{\mu\nu} (p_1,p_2)&=&
\frac{\dis N_c}{\dis 16 \pi^2} 
\varepsilon_{\mu\nu\beta\rho} p_1^\beta p_2^\rho 
\, F(p_1^2,p_2^2,q^2) \frac{\dis 2}{\dis M_i} \nonumber \\
{\rm with} \hspace*{1.5cm}
F(p_1^2,p_2^2,q^2)&=&1+I_3(p_1^2,p_2^2,q^2)-I_3(0,0,0)
\ea
where the form factor $I_3(p_1^2,p_2^2,q^2)$ is  the one given in 
appendix \ref{AppD}
and which appeared before in the study of the 
VPP three-point function  in section \rref{vppsub}. 
This form factor coincides with  the one 
found in the context of constituent quark-models (see for instance
\rcite{Ametller}) when the cut-off $\Lambda_\chi$ is sent to 
$\infty$. Here, this is a physical scale of
the order of the spontaneous symmetry breaking scale and 
therefore we have to keep it finite.
The anomalous Ward
identities in eq. \rref{anomW} is telling us that terms 
which are of chiral counting different to  
${\cal O}(p^4)$ have to be local chiral invariant \rcite{BP}
but they do not fix the regularization for those terms.
We therefore use  here
consistently the same regularization for them 
as in the non-anomalous sector. At ${\cal O}(p^4)$
the chiral anomaly also uniquely fixes the
 one-loop constituent chiral quark anomalous form factor
 to be the one in eq. \rref{onam} 
when $p_1^2=p_2^2=q^2=0$ \rcite{AB}.

Here we have used the anomalous Ward identity in  
eq. \rref{wardan}.
A naive use of the two-point functions and Ward identities 
would have led only
to the first term  in the sum in eq. \rref{PVVr1}. 
The second term is the result of
enforcing the validity of the QCD flavour anomaly.
Substituting the results on the two-point functions in section
\ref{twop} we  can write down the following explicit expression
\be
\begin{array}{l} \dis
\rlabel{PVVres}
\Pi^+_{\mu\nu}(p_1,p_2) = \frac{\dis N_c}{\dis 16 \pi^2} 
 \varepsilon_{\mu\nu\beta\rho} p_1^\beta p_2^\rho 
\left( \frac{\dis 4 M_i}{\dis g_S f_{ii}^2(-q^2)
 \left(m_{ii}^2(-q^2)-q^2\right)} \right) 
\nonumber \\ \dis
\left\{ 1 - g_A(-q^2)_{ii} \left[ 1 -
 F(p_1^2,p_2^2,q^2)\frac{\dis 
M_V^2(-p_1^2)_{ii} M_V^2(-p_2^2)_{ii}}{\dis \left(
M_V^2(-p_1^2)_{ii} - p_1^2\right) \left( M_V^2(-p_2^2)_{ii}
- p_2^2 \right)} \right] \right\}  \ . \nonumber \\
\end{array}
\ee
We refer to section \tref{threepvv}
for the definition of the physical anomalous 
$\pi^0\gamma^*\gamma^*$
 form factor after reducing this $PVV$ three-point function. 
We shall also discuss there on the VMD limit in this process and give some 
numerics.

\setcounter{equation}{0}
\section{Meson-Dominance}
\rlabel{VMD}

\subsection{Two-point functions}
Here we shall discuss the vector case, the axial-vector
case is similar. The transverse vector two-point function
in eq. \rref{transvec} reduces in the diagonal case, $m_i=m_j$
(the off-diagonal case can be done analogously) to the following
simple expression
\ba
\rlabel{twovec}
\Pi^{(1)}_V(-q^2) &=& 
2 f_V^2(-q^2)\, 
\frac{M_V^2(-q^2)}{M_V^2(-q^2)-q^2}  \, 
\\ {\rm with} \hspace*{0.5cm} \rlabel{fvmv2}
2 f_V^2(-q^2) M_V^2(-q^2) &=& \frac{N_c\Lambda_\chi^2}
{8\pi^2 G_V} \, \\ {\rm and} \hspace*{2cm}
2 f_V^2(-q^2) &=& {\overline{\Pi}^{(1)}_V(-q^2)}\, .
\ea
In the complete VMD limit this two-point function
has the same form but with $f_V$ and $M_V$ constants. Let
us see how complete VMD works in this model. For that, we shall
study the inverse of $\Pi^{(1)}_V(-q^2)$, which in the complete
VMD limit is a straight line. This function was plotted
in section \tref{numbers} in figure \tref{FigV1}. There we can see
that $\Pi^{(1)}_V(-q^2)$ in this model is very near of reproducing
the complete VMD linear form. Moreover, we can perform a linear
fit to the inverse of $\Pi^{(1)}_V(-q^2)$ to obtain the best VMD 
values for the $f_V$ and $M_V$ parameters. These parameters
are in this way meaningfully defined in the Euclidean region
$-q^2>0$ where the model is far from  the two constituent
quark threshold. Doing this type of fit for the values of 
the input parameters $\Lambda_\chi$, $G_V$, $G_S$ discussed
in section \tref{numbers} leads  
to $M_V \simeq 0.644$ GeV for the vector mass in the chiral limit
(remember that we are always in the large $N_c$ limit)
and $f_V \simeq 0.17$ for the decay constant.
For current quark masses values discussed also in section
\tref{numbers},  we obtain for the $\rho$ meson
flavour configuration $M_\rho \simeq 0.655$ GeV and
$f_\rho \simeq 0.17$ and  for the $\phi$ meson 
one $M_\phi \simeq 0.790$ GeV
and $f_\phi \simeq 0.14$. We see thus that the $\rho$ mass is
very close in the large $N_c$ limit,
to the one in the chiral limit, $M_V$.
Notice that these values for $M_V$ are far away from those quoted in
ref. \rcite{BBR}. The underlying reason is that 
 in ref. \rcite{BBR} $f_V$ and $M_V$ were determined
directly from the  Lagrangian at ${\cal O} (p^2)$ in the ENJL expansion,
identifying them with their values at $q^2=0$.
What we find here is that even though the
two-point function in eq. \rref{twovec} has the correct $q^2 \to
0$ limit behaviour it does have, with the choice of vector fields
to represent vector particles in ref. \rcite{BBR}, substantial
contributions from higher order terms (mainly of ${\cal O} (p^4)$
in the ENJL expansion). A physical 
vector field that would include these contributions can in
principle be defined as is shown by the fact that the inverse
of $\Pi^{(1)}_V(-q^2)$ is a rather straight line. What has happened
is that
\ba
\Pi^{(1)}_V(-q^2) &\simeq& \frac{\dis \left(2 f_V^2 M_V^2
\right)_{q^2=0}}{\dis M_V^2(0)-q^2 \left(1+\lambda+ 
{\cal O}(q^2/\Lambda_\chi^2)\right)} \, .
\ea
The vector meson mass derived in \rcite{BBR} was $M_V(0)$
while the slope of the physical two-point function 
(for $|q^2|/\Lambda_\chi^2<<1$ that is where this
ENJL cut-off model makes sense) corresponds to rather
$M_V \sim M_V(0)/\sqrt{1+\lambda}$. We find from the calculation
that indeed $\lambda$ is of order 1 ($\lambda \simeq 0.7$),
explaining the difference in the slope from the ${\cal O}
(p^2)$ ENJL calculation in ref. \rcite{BBR} 
of the two-point function to the ${\cal O}(p^4)$ one.

We can also see from eqs. \rref{46}, \rref{pipa0}-\rref{pip} 
and \rref{pisca} that the forms of these two-point functions
are very similar to the corresponding ones
 in the meson dominance limit 
but with couplings varying with $q^2$. The identification
of the corresponding physical values will involve analogous
procedures to the one described above for the transverse 
vector two-point function one.

\subsection{VPP three-point function}
\rlabel{threevpp}

In this subsection we discuss how the result for the three-point
function $\Pi^{VPP}_\mu(p_1,p_2)$ obtained in section
\tref{vppsub} can be used to determine the physical
pion electromagnetic form factor in this model.
We shall discuss the $VPP$ three-point function flavour
structure corresponding to the 
three-point function $\Pi^+_{\mu}(p_1,p_2)$ 
in eq. \rref{VPPres} for $m \equiv m_i=m_j=m_k$ 
and $p^2 \equiv p_1^2=p_2^2$ for definiteness.

Since this $\Pi^+_{\mu}(p_1,p_2)$ 
is a Green's function we first have to reduce
the external legs to properly normalized pion fields.
The vector leg acts here as an external source and is properly
reduced without bringing in any factor.
For this, we first look at the pseudoscalar two-point function
in eq. \rref{pip} obtained using the same external fields
and parametrize it around the pole as
\be \rlabel{pip2}
\Pi_P(-p^2)=-\frac{\dis 1}{\dis g_S} +
\frac{\dis Z_\pi}{\dis p^2- m_\pi^2} \left( 1 +
{\cal O}(m_\pi^2/\Lambda_\chi^2) \right) \, .
\ee
The reducing factor $Z_\pi$ is
\ba
\rlabel{zpi}
Z_\pi &\equiv&-\frac{\dis \left( M_i +M_j \right)^2}{\dis 2 f_\pi^2
(-m_\pi^2) g_S^2} \, \frac{\dis 1}{\dis A^2}\, \hspace*{1.5cm} 
{\rm with}  \nonumber \\ \rlabel{A2}
A^2&=& 1 - \left. \frac{\dis \partial m_{ij}^2(-p^2)}
{\dis \partial p^2} \right|_{p^2=m_\pi^2}
\, \nonumber \\ 
&=& 1+ \frac{\dis g_A^2(-m_\pi^2)}{\dis 2 f_\pi^2(-m_\pi^2)}
\left[\overline f^2_\pi(0) - \overline f^2_\pi(-m_\pi^2)
+ 2 m_\pi^2 I_3(m_\pi^2,m_\pi^2,0) \right] \nonumber \\ 
\ea
where $I_3(p_1^2,p_2^2,q^2)$ defined in appendix \tref{AppD}
and $f_\pi^2(-q^2)$ and  ${\overline f}_\pi^2(-q^2)$
in eqs. \rref{104}-\rref{105}.
The quantity $A$ is very close to one and exactly one in the chiral limit. 
Each pion leg brings a factor
$Z_\pi^{1/2}$ after reducing the Green's function
to the physical amplitude. 
Rewriting the pseudoscalar two-point function in the form
in eq. \rref{pip2} gives that $m_\pi^2$ is the
solution of $ m_\pi^2= m_{ij}^2(-m_\pi^2)$.

Reducing the $VPP$ three-point function $\Pi^+_{\mu}(p_1,p_2)$
in eq. \rref{VPPres} we find
that it can be written as follows\footnote{To obtain the $\gamma^*\pi^+\pi^-$
three-point function from this $\Pi^+_\mu$ is necessary to
multiply it by the electric charge of the pion.}
(we shall suppress the flavour indices 
which are always $ii$)
\be
\renewcommand{\arraystretch}{1.5}
\begin{array}{l}
\Pi^{+\mu}(p_1,p_2) = \frac{\dis Z_\pi}{\dis (p^2-m_\pi^2)^2} 
\, F_{VPP}(p^2,q^2) \, (p_2-p_1)^\mu \,   
\nonumber \\
\end{array}
\renewcommand{\arraystretch}{1}
\ee
which defines  the electromagnetic pion form factor 
(or in general the pseudoscalar vector form factor)
$F_{VPP}(p^2,q^2)$  in this model. (The
general pion form factor, i.e different quark masses and
$p_1^2 \ne p_2^2$ can be obtained similarly from 
\rref{VPPres}.) This form factor in the ENJL model 
is  expected to be a good approximation 
at intermediate and low-energy energies, 
within the validity of the ENJL
model we are working with, i.e. for $|q^2| << \Lambda_\chi^2$.
The explicit expression for this form factor\footnote
{This form factor
was also calculated in ref. \rcite{Meissner}. With
the appropriate changes of notation it agrees 
with the one found there.}  is
\ba
\rlabel{fvpp2}
F_{VPP}(m_\pi^2,q^2) &=& \frac{\dis 1}{\dis 2 A^2 f_\pi^2
(-m_\pi^2)} \, \frac{\dis M_V^2(-q^2)}{\dis M_V^2(q^2)-q^2}
\left\{ 2 f_\pi^2(-m_\pi^2) \right. \nonumber \\
&-&  q^2 (1-g_A(-m_\pi^2))^2
f_V^2(-q^2)  +  \frac{\dis 2 g_A^2(-m_\pi^2)}{\dis
q^2 -4m_\pi^2} \nonumber \\ &\times& \left.
\left[(q^2-2 m_\pi^2)(\overline f^2_\pi(-q^2)
- \overline f^2_\pi(-m_\pi^2)) - 4 m_\pi^4 I_3(m_\pi^2,
m_\pi^2,q^2) \right] \right\} \, . \nonumber \\
\ea
Notice that this form factor has  no pole at $q^2=4m_\pi^2$.
The value of $A^2$ in eq. \rref{A2} is precisely the one 
that ensures that $F_{VPP}(m_\pi^2,0)=1$ in the large $N_c$ 
limit as is required by the electromagnetic gauge invariance.
This must be so since we have imposed the Ward identities
to obtain this form factor.
In figure \ref{figVPP} we have plotted the inverse of this
form factor for the parameters quoted in section \tref{numbers}
in the chiral case ($\overline m=0$) 
and in the case corresponding to the 
physical pion mass ($\overline m=3.2$ MeV).
\begin{figure}
\rotate[r]{\epsfysize=13.5cm\epsfxsize=8cm\epsfbox{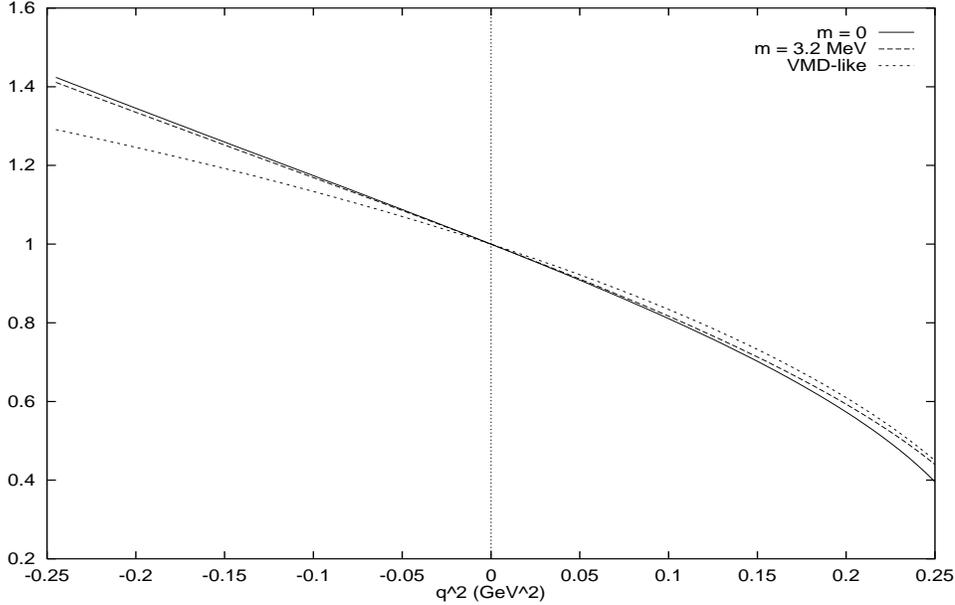}}
\caption{The inverse of the 
vector form factor of the pion of eq. \protect{\rref{fvpp2}}.
For the chiral limit and with all current quark masses 
equal to 3.2 MeV. Also plotted is the VMD approximation 
$M_V^2(-q^2)/(M_V^2(-q^2)-q^2)$ for the latter case.}
\rlabel{figVPP}
\end{figure} 
As can be seen from the picture, it is a rather straight line so 
the complete VMD result for this form factor, i.e.,
\ba
\rlabel{vppvmd}
F_{VPP}^{VMD}(m_\pi^2,q^2) =
\frac{\dis M_\rho^2}{\dis M_\rho^2 - q^2} 
\ea
with constant vector mass $M_\rho$ works rather well.
The slope of the linear fit of the inverse of the form factor
in eq. \rref{fvpp2} to this VMD form
gives a vector mass which is 
$M_\rho \simeq 0.77$ GeV. This mass is 
very close to the physical value and rather
different from the one found for the transverse vector 
two-point function in the VMD limit $M_\rho \simeq
0.655$ GeV in the large $N_c$ limit. 
This explains why using the physical
$\rho$ meson mass and the VMD dominance works so well 
but it also shows that this $M_\rho$ ``mass'' in eq. 
\rref{vppvmd} has not, in principle, to be the 
same as the mass of the vector meson described
by the transverse two-point vector function. 

The same three-point function $VPP$ also
contains implicitly 
the $\rho \to \pi \pi$ coupling constant $g_V$.
(See ref. \rcite{BBR} for its definition. Notice that is
 different from the symbol used in section \tref{twop}.) 
Again, to obtain
the physical $\rho \to \pi \pi$ amplitude
we should first reduce the vector leg that now corresponds
to the $\rho$ particle, (remember that the pion legs have
been already reduced). This will bring a factor which
is similar to the factor 
$1+\lambda$ discussed in the previous subsection.
We shall, as before, first determine the reducing vector
factor from the vector two-point function in eq. 
\rref{twovec}. The reducing factor $Z_\rho$ is 
\ba
\rlabel{zrho}
Z_\rho &\equiv& - \left(\dis 2 f_V^2 M_V^2\right)
\left(1- \left. \frac{\dis \partial M^2_V(-q^2)}{\dis
\partial q^2}\right|_{q^2=M_\rho^2} \right)^{-1} 
\nonumber \\
&\equiv& - \frac{\dis \dis 2 f_V^2 M_V^2}{\dis B^2}  \, .
\ea
In this equation the combination $2f_V^2M_V^2$ is the one given
in eq. \rref{fvmv2} and is independent of $q^2$.
The vector mass $M_\rho$ is again given by the solution to
$M_\rho^2= M_V^2(-M_\rho^2)$. 

One also can rewrite down the electromagnetic
pion form factor showing explicitly the coupling constant
of the $\rho$ meson to pions, $g_V$, as follows
\ba
\rlabel{defvgv}
F_{VPP}= 1 + f_V g_V \, 
\frac{\dis q^2}{\dis f_\pi^2} \frac{\dis M_\rho^2}{\dis
M_\rho^2 -q^2} \,  .
\ea
Then, in the complete VMD limit one has $f_V g_V = f_\pi^2/M_\rho^2$.
In this ENJL model this relation is equivalent
to $g_V = (1-g_A) f_V$, i.e. one has complete
VMD and the KSRF relation \rcite{KSRF}
$2 g_V = f_V$ satisfied for $g_A=1/2$.

One can see in the eq. \rref{defvgv}, that 
reducing the $\rho$ vector leg brings in a factor $B^2$ in the
numerator and another factor $B^2$ in the denominator with
the net result that $f_V(-q^2)g_V(-q^2)$ is not affected
by  reducing of the vector leg as much as  happens to 
$f_V^2(-q^2)$ in eq. \rref{twovec}.
Then, with the definition of $f_V g_V$ in eq. \rref{defvgv}
we get the following
\be
\renewcommand{\arraystretch}{1.5}
\begin{array}{l}
\rlabel{KSR} \dis
f_V(-q^2) g_V(-q^2)  \equiv \nonumber \\
\dis \frac{\dis 1}{\dis 2 A^2 q^2}
\left[ \left(q^2 - M_V^2(-q^2)\right)
\left(2 A^2 f_\pi^2(-m_\pi^2)/M_V^2(-q^2) - f_V^2(-q^2) 
(1-g_A(-m_\pi^2))^2\right) \right. \nonumber \\
\dis + (1+g_A(-m_\pi^2)) f_\pi^2(-m_\pi^2) + \frac{\dis 2 
g_A^2(-m_\pi^2)}{\dis q^2 -4m_\pi^2} \nonumber 
\\ \dis \left. \times \left( (q^2-2m_\pi^2)
({\overline f}_\pi^2(-q^2) - {\overline f}_\pi^2(-m_\pi^2))
-4 m_\pi^4 I_3(m_\pi^2,m_\pi^2,q^2) 
\right) \right] \,  .  \nonumber \\
\end{array} 
\renewcommand{\arraystretch}{1}
\ee
Notice that this $f_V(-q^2)g_V(-q^2)$ form factor
has neither a pole at $q^2=0$ nor at $q^2=4m_\pi^2$ and 
when expanded in $q^2$ and with $m_\pi^2=0$
one gets $f_V(0) g_V(0) = 2 L_9$, where $L_9$
 is the one found in ENJL in ref. \rcite{BBR}.
As discussed there, at $q^2=0$ 
one has the KSRF \rcite{KSRF} relation,
i.e. $f_V(0) = 2 g_V(0)$ (which is valid for 
$q^2=m_\pi^2=0$)
analytically for $g_A=0$ and very approximately 
satisfied for $g_A$ varying between 0 and 1.
The expression in eq. \rref{KSR} is the off-shell
equivalent to the KSRF relation in this model. For 
$g_A=0$ the vector mass vanishes and the $\rho$ meson
couples as an SU(3)$_V$  gauge boson, in fact in this limit
one recovers the results of the Hidden Gauge Symmetry
model \rcite{HGS} for the non-anomalous sector.
In particular, when $g_A =0$ we have that the reducing
factor $B$ is 1
as corresponds to external gauge sources. In this limit
($g_A=0$),
one still has the KSRF relation analytically satisfied
off-shell, i.e. $f_V(-q^2)= 2 g_V(-q^2)$ for all
$q^2$. 

In the limit $g_A \to 1$ one has 
\be
\renewcommand{\arraystretch}{1.5}
\begin{array}{l}
\rlabel{KSR2} \dis
f_V(-q^2) g_V(-q^2)  \to \dis \frac{\dis 1}{\dis A^2 q^2}
\left[ f_\pi^2(-m_\pi^2) (1-A^2)  + 
\frac{\dis 1}{\dis q^2 -4m_\pi^2} \right. \nonumber 
\\ \dis \left. \times \left( (q^2-2m_\pi^2)
(f_\pi^2(-q^2) - f_\pi^2(-m_\pi^2))
-4 m_\pi^4 I_3(m_\pi^2,m_\pi^2,q^2) 
\right) \right] \,  .  \nonumber \\
\end{array} 
\renewcommand{\arraystretch}{1}
\ee
This is the constituent quark model result (in $g_A=1$
the vector mesons decouple from this model) and when expanded
in $q^2$ with $m_\pi^2=0$ it coincides with the corresponding
result in ref. \rcite{BBR}. The KSRF relation is 
not analytically fulfilled in this limit but
on can see that analytically is very approximately
satisfied. Then, we have that for $g_A$ 
varying between 0 (the gauge vector limit) and 1 (the 
constituent quark limit) the KSRF relation goes from
being analytically fulfilled to be very approximately
fulfilled for any value of $q^2$.

Let us see how $g_V(-q^2)$ works numerically
compared with $f_V(-q^2)$ for a definite value of $g_A$. 
In figure \ref{fvgvfig} 
we plot $f_V(-q^2)/2$ and $g_V(-q^2)$ for the values of parameters
discussed in section \tref{numbers}. These values 
correspond to $g_A(0)=0.61$.
\begin{figure}
\rotate[r]{\epsfysize=13.5cm\epsfxsize=8cm\epsfbox{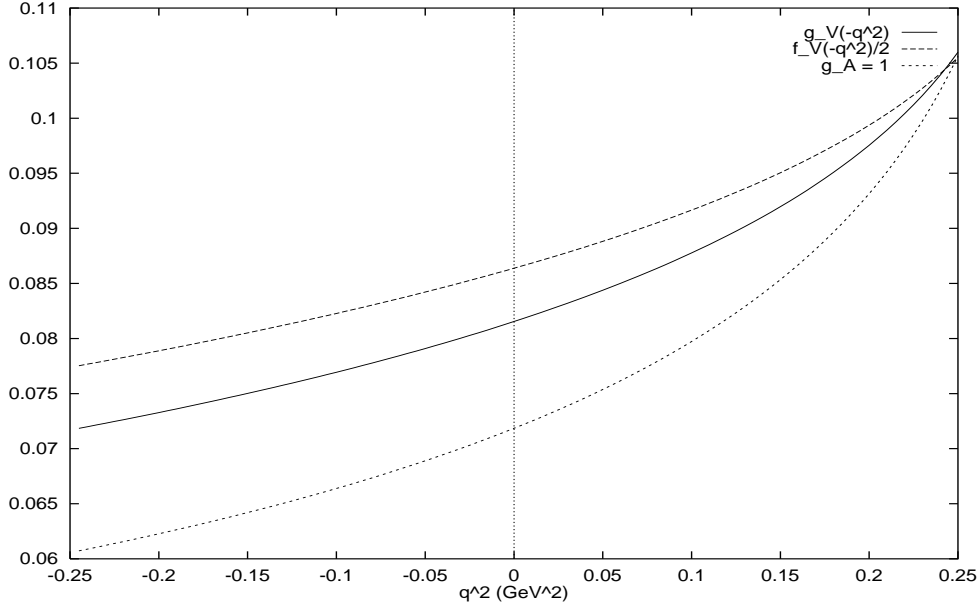}}
\caption{The generalized KSRF relation. We plot $g_V(-q^2)$ for
$g_A =0.61$ (solid line); $g_A \to 1$ (short-dashed line)
and $f_V(-q^2)/2$ (dashed line).
 The difference between the curves gives the violation
of the KSRF relation. See text for further comments.}
\rlabel{fvgvfig}
\end{figure}  
The form factor $g_V(-q^2)$ is somewhat dependent
on $q^2$ with $ (2.1\sim 2.2) \, 
g_V(-q^2) \simeq f_V(-q^2)$ in the Euclidean region.
In this figure we also plot the case $g_A \to 1$
where the same features can be seen. The form factor
$g_V(-q^2)$ for any value of
$g_A$ will be between the line $f_V/2$ (i.e., the $g_A=0$
limit) and the line for $g_A=1$, therefore the KSRF relation
is approximately satisfied off-shell for any value of $g_A$.

\subsection{PVV three-point function}
\rlabel{threepvv}

In this subsection we want to study the $\pi^0\gamma^*\gamma^*$
anomalous form factor. For that we shall reduce 
the $PVV$ Green's function in eq. \rref{PVVres} calculated in 
section \tref{pi0gg} to the physical amplitude following the
same procedure that in the previous section (for details
see there). Now, we have to reduce one pion leg, this will
bring in a factor $\sqrt Z_\pi$ and two external vector sources
legs which are properly reduced without bringing any factor.
Then the $PVV$ three-point function in eq. \rref{PVVres}
\footnote{To obtain the $\pi^0\gamma^*\gamma^*$  three-point 
function from this $\Pi^+_{\mu\nu}$ is necessary to multiply it by
a factor $\sqrt 2$ coming from the $\pi^0$ flavour structure
and a factor $e^2/3$ from the quarks electric charge.}
can be rewritten as follows
\ba
\Pi^+_{\mu\nu}(p_1,p_2) &=& \frac{\dis \sqrt Z_\pi}{\dis
q^2-m_\pi^2} \, 
\frac{\dis N_c}{\dis 16\pi^2} i \varepsilon_{\mu\nu\beta\rho}
\, p^\beta_1 p^\rho_2 \, 
\frac{\dis 2 \sqrt 2 }{\dis f_\pi(-m_\pi^2)}
\nonumber \\
&\hspace*{1.5cm} \times&  F_{PVV}(q^2,p_1^2,p_2^2) \,  
\ea
where $F_{PVV}$ is the $\pi^0 \to \gamma^* \gamma^*$
form factor in this model.
Notice that the reducing factor $A$ in eq.
\rref{A2} goes to one in the chiral limit preserving,
in that way, the chiral anomaly condition $F_{PVV}(0,0,0)=1$.
This form factor can be used as an accurate
interpolating expression in low-energy hadronic processes
valid for external momenta smaller than $\Lambda_\chi^2$.
We plot the inverse of
this form factor for the case $p_2^2=0$ in figure
\ref{pvvfig}. 
\begin{figure}
\rotate[r]{\epsfysize=13.5cm\epsfxsize=8cm\epsfbox{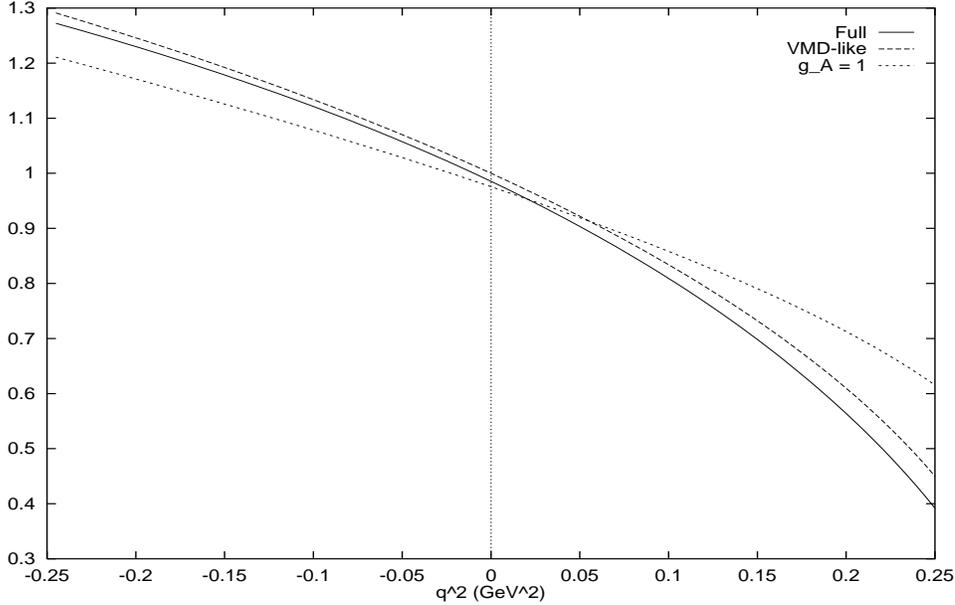}}
\caption{The 
inverse of the $\pi^0\gamma^*\gamma$ form factor for one photon on-shell and
one off-shell as a function of the photon mass squared, $q^2$. Notice the
linearity in the Euclidean region. Plotted are the full result,
$M_V^2(-q^2)/(M_V^2(-q^2)-q^2)$(VMD-like) and the ENJL model without
vector and axial-vector mesons ($g_A = 1$).}
\rlabel{pvvfig}
\end{figure}
Notice that there $F_{PVV}(m_\pi^2,0,0) \neq 1$
and the difference comes from the reducing factor $A$
and is of chiral counting ${\cal O}(p^6)$.
We can expand this form factor for small $p_1^2,p_2^2$
and pion mass \footnote{For the $\pi^0$ decay we are on the pole
and hence $q^2=m_\pi^2$} as follows
\ba
F_{PVV} (m_\pi^2,p_1^2,p_2^2)
 &=& 1 + \rho \, (p_1^2+p_2^2)  + \rho' \, 
m_\pi^2 + {\cal O} (q^4) \, ,
\ea
this expansion defines the slopes $\rho$ and $\rho'$
which in this model are
\ba
\rho &=& g_A(0) \left(\frac{\dis 1}
{\dis M_V^2(0)} + \frac{\dis \Gamma(2,M^2/\Lambda_\chi^2)}
{\dis 12 M^2} \right) \, ,
\nonumber \\ {\rm and} \hspace*{1cm} 
\rho' &=& g_A(0) \left(\frac{\dis \Gamma(2,M^2/\Lambda_\chi^2)}
{\dis 12M^2}-\frac{\dis \Gamma(1,M^2/\Lambda_\chi^2)}
{\dis 12 \Gamma(0,M^2/\Lambda_\chi^2) M^2}
\right)\, .
\ea
Where the second term in $\rho'$ comes from the reducing
factor $A$ defined.
The constituent quark mass $M$ here
is the one corresponding to the current quark mass value
$\overline m = 3.2$ MeV
used in the numerical applications section \tref{numbers}.
Using $M_V^2(0)=6 M^2 g_A(0) / (1-g_A(0))$ \rcite{BBR}
we can write down them as  
\ba
\rlabel{slope}
\rho &=& \frac{\dis 1}
{\dis 12 M^2}  \left(2 - 
\left(2-\Gamma(2,M^2/\Lambda_\chi^2)\right) g_A(0)\right) 
\, \nonumber \\
{\rm and} \hspace*{1cm} && \nonumber \\
\rho' &=& \frac{\dis g_A(0)}
{\dis 12 M^2} \left(\Gamma(2,M^2/\Lambda_\chi^2)
- \frac{\dis \Gamma(1,M^2/\Lambda_\chi^2)}
{\dis \Gamma(0,M^2/\Lambda_\chi^2)} \right) \, \nonumber \\
\ea
which interpolate between the constituent quark-model
result $g_A(0)=1$ and the gauge vector meson result $g_A(0)=0$.

With the input parameters we have been using 
(see numerical application section \tref{numbers})
we get
\ba
\rho=(0.86+0.67)=1.53 \, {\rm GeV}^{-2} \, 
\nonumber \\ {\rm and} \hspace*{1cm}
\rho'=(0.67-0.27)=0.40 \, {\rm GeV}^{-2} \, .
\ea
Where for $\rho$ the first number between brackets
is the vector meson exchange contribution and the second
is the constituent quark contribution (up to $g_A(0)$). We see that
both contributions are very similar  giving some
kind of complementarity between both approaches and 
explaining the relative success of both when 
used to describe this slope. For $\rho'$ they are the constituent quark 
contribution and the one coming from the 
pion leg reducing factor $1/A$. (Notice the cancellation
there.)

In the limits $g_A \to 1$  and $g_A \to 0$  we find
\ba
\rho=1.10 \, {\rm GeV}^{-2} &{\rm for}& g_A \to 1 \, , 
\nonumber \\ 
\rho=2.20 \, {\rm GeV}^{-2} &{\rm for}& g_A \to 0 \, ,
\nonumber \\
\rho'= 0.66 \, {\rm GeV}^{-2} &{\rm for}& g_A \to 1 \, , 
\nonumber \\ {\rm and} \hspace*{1cm}
\rho'= 0.00 \, {\rm GeV}^{-2} &{\rm for}& g_A \to 0 \, .
\ea 
We see that the difference between these two limits
is big and that the actual result is some kind of
interpolation. Experimentally \rcite{CELLO}
\ba
\rho=(1.8 \pm 0.14) \, {\rm GeV}^{-2} \, .
\ea
Taking into account that the $1/N_c$ 
corrections from $\chi$PT loops are estimated \rcite{Hans}
to be twice the experimental error 
we consider the result as good.

Let us compare this full result in eq. \rref{slope}
with the one obtained in ref. \rcite{Ximo} in this same
model assuming complete VMD in the chiral limit. 
There, the same prescription 
to include the QCD chiral anomaly that here \rcite{BP}
was used at the one-loop level with the result
\ba
\rlabel{slope0}
\rho &=& \frac{\dis 1}{\dis 12 M^2} \, 
\frac{\dis 1-g_A^2(0)}{\dis g_A(0)} \, . 
\ea
Of course, this complete VMD result vanishes when $g_A=1$
where vector mesons decouple.
The differences between eq. \rref{slope} and eq. 
\rref{slope0} come from the resummation of all the
orders in external momenta that are included in the full
result in eq. \rref{slope}. One can see that the complete 
VMD result coincides with the full result for $g_A(0)
\simeq0.50$.

\setcounter{equation}{0}
\section{Conclusions}
\rlabel{conclusions}
In this paper we have derived in a general class of ENJL-like models
the two-point functions away for the chiral limit and for different masses
in terms of the one-loop ones. This derivation used the Ward identities
of the one-loop functions. The heat kernel expansion yields two more identities
then can be derived from the current identities directly. These are then
used to rewrite all two-point functions in terms of two basic ones. These
can then be calculated in the bare ENJL-model as we did here or gluonic 
background corrections can be taken into account (see ref. \rcite{BRZ}
for a discussion). These extra identities allow us to discuss the 
Weinberg Sum Rules in this class of models. 
We find that the first WSR is satisfied and the third one is
broken in the same way as required in QCD. The second WSR
 is satisfied in this model while QCD requires it to be broken.
The very high energy behaviour is thus a little too suppressed. 

We also find (in the case of equal masses)
the simple formula for the scalar mass that had been argued before when only
divergent terms were kept in the heat kernel expansion (see second paper
in ref. \rcite{ENJL}). All two-point functions can also be written in
a form very like a form of meson dominance but with the couplings and masses
depending on the momentum. In the Euclidean region, $q^2\le0$, the vector
two-point function can also be well described by a VMD form with constant 
couplings. The relation of these with those from the low-energy expansion
was treated as well.

We then proceeded to calculate two examples of three-point functions. Again
the use of Ward identities simplified the calculation and pinpoints all the
regularization ambiguities into the one-loop function. We want to point 
out that both the anomalous sector and the non-anomalous sectors
of these ENJL-like models are then treated on the same foot and VMD
can discussed in both sectors with a unique prescription, namely the
use of the relevant Ward identities \rcite{BRZ,BP}. Then
the regularization dependence uncertainties are consistently
treated within the same prescription in
both the anomalous sector and the non-anomalous sector.
Here we discussed
Vector Meson Dominance and the KSRF relation for the VPP case.
We then use the Ward identities, modified
to reproduce the QCD flavour anomaly, to calculate the PVV function.
Here we find that naive VMD expectations for this function cannot
be realized. No simple generalization to $q^2$ dependent couplings is
possible.
Formally our expression looks very much like the VMD expression
with couplings and vector mass running with $q^2$
times $g_A(-q^2)$ plus a second term coming from the requirements of
the anomaly. The numerical result for the slope is, however, in good agreement
with the VMD value. But the $g_A(-q^2)$ factor diminished the ``real'' VMD
part to a little more than half the full value while the constituent quark
loop adds the remainder. Here we reconcile both explanations for the slope,
the VMD one and the quark loop one, \rcite{Ametller,Hans}.

\section*{Acknowledgements}
We would like to thank E. de Rafael for encouragement.
The work of JP has been supported in part by 
 CICYT (Spain) under Grant Nr. AEN93-0234.

\appendix
\def\theequation{\Alph{section}.\arabic{equation}}
\setcounter{equation}{0}
\section{Proper time and incomplete Gamma functions}
\rlabel{GAMMA}

In this appendix we give the regularization method we have used
throughout this work and some related definitions.
After performing the standard Feynman parametrization,
 one constituent fermion propagator is regulated
consistently in this work 
using a proper time regulator as follows
\ba
\frac{\dis 1}{\dis M^2(Q^2,x)} &\Rightarrow&
{\dis \int^\infty_{1/\Lambda_\chi^2}} {\rm d}\tau
\, e^{-\tau M^2(Q^2,x)} \, .
\ea
After performing the remaining $Q^2$ integration 
and the change of variables $\tau \Lambda^2_\chi \to z$
one arrives to the following type of integrals
\ba
\Gamma \left(n-2,\epsilon \right) &=&
{\dis \int^\infty_\epsilon}
\frac{\dis {\rm d}z}{\dis z} \, z^{n-2} e^{-z} \, ,
\ea
with $\epsilon \equiv M^2(Q^2,x)/\Lambda_\chi^2$ and 
$n=1,2, \cdots$. These $\Gamma(m,y)$ are the so-called
incomplete Gamma functions.

In general, this regulator breaks the Ward identities.
We have, however, always imposed all the Ward identities
explicitly so our results have the correct symmetry
covariance.

\setcounter{equation}{0}
\section{Derivation of the Ward identities}
\rlabel{AppB}

In this appendix we generalize the proof in the appendix of ref. \rcite{BRZ}
to the case with nonzero current quark masses. There a proof was given
of all relevant identities in terms of the heat kernel expansion 
(for an excellent
 recent review and definitions see ref. \rcite{Ball2}) and some
of them in terms of the Ward identities as well. Here those which can
be derived directly from the Ward identities can also be derived from the
heat kernel expansion but since they involve different masses they
require a resummation of different terms. For these the direct derivation
of the Ward identities is actually simpler. Only for the additional relations
will we give the heat kernel derivation.

At the one-loop level we use as Lagrangian the one in
eq. \rref{ENJL} (with the same definitions as there)
\be
\rlabel{lagrang}
{\cal L}_{\rm ENJL} = \overline{q}{\cal D}{q}
\ee
where $\cal D$ contains the couplings to the external 
fields $l_\mu,r_\mu,s$
and $p$ as well as the effects of the four-quark terms in 
${\cal L}^{\rm S,P}_{\rm NJL}$ and ${\cal L}^{\rm V,A}_{\rm
 NJL}$ on the quark currents at the one-loop level.
In particular it contains the constituent quark masses, $M_i$.
However, we shall keep the notation $l_\mu,r_\mu,s$ and $p$ 
to denote the quark current sources in the presence of these
four-quark NJL operators.
The one-loop current identities derived from this Lagrangian are
\ba
\rlabel{curr}
\partial^\mu V^{ij}_\mu &=& -i\left(M_i - M_j\right) S^{ij}
\nonumber\\
\partial^\mu A^{ij}_\mu &=& \left(M_i + M_j\right) P^{ij}\ .
\ea
When the whole series of constituent quark bubbles are summed
these identities are satisfied changing constituent quark masses
by current quark masses.
In addition we use the equal time commutation relations for fermions
\be
\rlabel{eqtime}
\left\{ q^{i\dag}_\alpha(x) , q^j_{\beta}(y)\right\}_{x^0 = y^0}
=
i\delta_{\alpha\beta}\delta_{ij}\delta^3({\bf x} - {\bf y})\ .
\ee
Here $\alpha$ and $\beta$ are Dirac indices and $\bf x$ means the spatial
components of $x$.
Multiplying the two-point functions with $i q_\mu$ 
is equivalent to
taking a derivative of the exponential under the integrals in eqs.
\rref{9} to \rref{12}. By partial integration we then get several terms,
those due to the time ordering which leads to equal time commutators
and those where the derivative hits one of the currents. The first type
are evaluated using eq. \rref{eqtime} and the second type are related to
other two-point functions using eq. \rref{curr}. This then leads
to the expressions \rref{ward1} to \rref{ward2}.

The derivation of the other two identities is slightly more complicated.
The effective action of the Lagrangian in eq. \rref{lagrang} can
be obtained in Euclidean space as a heat kernel 
expansion (see ref. \rcite{Ball2}). 
The coefficients of this expansion are 
the so-called Seeley-DeWit coefficients, they
are constructed out of the two quantities $E$ and $R_{\mu\nu}$.
These are defined as
\ba
\rlabel{ER}
{\cal D}^\dagger {\cal D} &\equiv& 
-\nabla_\mu\nabla^\mu + E +\overline{M}^2\ ,
\nonumber\\
R_{\mu\nu} &\equiv& \left[\nabla_\mu ,\nabla_\nu\right]\ ,
\nonumber \\
\nabla_\mu \# &\equiv& \partial_\mu \# -i [v_\mu,\#] 
-i [a_\mu \gamma_5 ,\#]\, .
\ea
If in eq. \rref{lagrang} the Dirac operator $\cal D$ contains couplings to
gluons these should not be taken into account in eq. \rref{ER}. The
relevant heat kernel expansion in that case will have different coefficients
depending on vacuum expectation values of gluonic operators, but will
still be constructed out of the quantities in eq. \rref{ER}
(depending now also on the gluon field). The quantity
$\overline{M}$ is the mass that is used in the heat kernel 
expansion. The operator $\cal D$ is 
\be
i\gamma^\mu(\partial_\mu -i v_\mu -i a_\mu\gamma_5)
-{\cal M}-s+ip\gamma_5\ .
\ee
Here ${\cal M}={\rm diag}(m_u,m_d,m_s)$ is the current quark mass
matrix and we allow for spontaneous chiral symmetry breaking
solution $\langle 0 | s(x) | 0 \rangle \neq 0$.
For the terms relevant to two-point functions we
have 
\ba \rlabel{E}
R_{\mu\nu} &=&-i(v_{\mu\nu}+a_{\mu\nu} \gamma_5) \, , \nonumber 
\\ {\rm and} \hspace*{2cm} && \nonumber \\
E &=& -\frac{i}{2}\sigma^{\mu\nu}R_{\mu\nu}
+i\gamma^{\mu} {\rm d}_\mu \left(M + s + i p\gamma_5 \right)
\nonumber \\ &-& \gamma^\mu
\left\{a_\mu \gamma_5,M+s-ip\gamma_5 \right\} +
\left\{M,s\right\}-i\left[M,p\right]\gamma_5
+s^2+p^2 
\nonumber\\&+&
 M^2 -\overline{M}^2\ , \nonumber \\
{\rm with} && \nonumber \\
v^{\mu\nu} &\equiv& \partial^\mu v^\nu - \partial^\nu v^\mu
-i [v^\mu,v^\nu] \, , \nonumber \\
a^{\mu\nu} &\equiv& \partial^\mu a^\nu - \partial^\nu a^\mu
-i [a^\mu,a^\nu] \, , \nonumber \\
{\rm d}^\mu \# &\equiv& \partial^\mu \# - i [v^\mu,\#] \, .
\ea
The main difference with ref. \rcite{BRZ} is the occurrence 
of the last line in the expression for $E$ in
\rref{E}. We shall call this last line $E_0$.
In this equation, $M \equiv 
{\rm diag}(M_u,M_d,M_s)$, 
the diagonal matrix of the constituent quark masses
defined in eq. \rref{gap}. Notice that the scalar field here has
been shifted and we have now $\langle 0 | s(x) | 0 \rangle =0$
(though we use the same notation for it). 
When $G_S \to 0$ in eq. \rref{gap}
then ${\overline M} \to 0$ and $M \to {\cal M}$.
 Let us now 
systematically go through
all possible types of terms in the expansion. We shall not discuss the mixed
two-point functions here since we only want to prove eqs. 
\rref{ward5}-\rref{ward6}.

In the heat kernel expansion, those terms 
containing two factors  $R_{\mu\nu}$ only contribute to the
transverse parts, $\ovpi^{(1)}_{V,A}$ and in the same way. Their contributions
hence obviously satisfy eqs. \rref{ward5}-\rref{ward6}. Similarly, 
one factor $R_{\mu\nu}$ requires the presence of two covariant 
derivatives $\nabla_\mu$. 
By commuting derivatives (the extra terms only contribute
to three and higher point functions) 
and partial integration these can be brought next to each other so they convert
into a second factor  $R_{\mu\nu}$. This brings us back to the
previous case. Intervening $E$'s can only contribute via $E_0$ but these
do not spoil the above argument.
The first term in $E$, namely $\sigma_{\mu\nu}R^{\mu\nu}$, 
requires a 2nd $\sigma_{\mu\nu}
R^{\mu\nu}$ because otherwise the trace over Dirac indices vanishes.
These also behave like terms with two factors $R^{\mu\nu}$. Therefore,
in the remainder we are only concerned with $E$ without this first term.

$E$ can also directly contribute to the scalar and pseudoscalar two-point
function in the same way via $s^2 + p^2$. Extra factors $E$ become again
$E_0$ and extra derivatives also respect the relation \rref{ward6}.
The most complicated case is where both fields come from a different $E$.
This contributes in the form $E_0^n E E_0^m \partial^{2i} E$.
These contribute to all form factors in the form $M_i^n M_j^m q^{2i}$
times the coefficients listed in Table \tref{table1}.
\begin{table}[htb]
\begin{center}
\begin{tabular}{|c|c|}
\hline
Function & Contribution \\
\hline
$\ovpi_S $&$ -q^2+(M_i+M_j)^2 $\\
$\ovpi_P $&$ -q^2+(M_i-M_j)^2 $\\
$\ovpi^{(0)}_A $&$ (M_i+M_j)^2/q^2$\\
$\ovpi^{(1)}_A $&$ -(M_i+M_j)^2/q^2$\\
$\ovpi^{(0)}_V $&$ (M_i-M_j)^2/q^2$\\
$\ovpi^{(1)}_V $&$ -(M_i-M_j)^2/q^2$\\
\hline
\end{tabular}
\end{center}
\caption{The contribution of terms of the type $E^{m+n+2}$ to the two-point
functions.\rlabel{table1}}
\end{table}
These coefficients obviously satisfy the relations 
\rref{ward5}-\rref{ward6}. 
The last type of terms is where  the external fields come out of a derivative.
We do not consider the mixed case here, so both the fields have to come
out of a derivative due to the $\gamma_\mu$ that is necessarily present
in the $E$ that would be a candidate for the external field.
So there are those where the external fields are contained in
two factors $\nabla_\mu$. 
If the indices of these are different, then there need to be
at least two extra derivatives present that will produce a $q_\mu q_\nu$.
This contributes equally to $\ovpi^{(0)}_V$ and $\ovpi^{(0)}_A$.
If the indices are equal, it will contribute proportional to $g_{\mu\nu}$
and thus to the vector and axial-vector equally with
$\ovpi^{(0+1)}_{V,A} = 0$. This completes the proof of the identities
\rref{ward5}-\rref{ward6}.

Now it remains to prove that these contributions will never produce a
pole in $\ovpi^{(0+1)}$ at $q^2=0$. 
Terms that contain two factors $R_{\mu\nu}$ contain 
two factors of momenta and hence do not. Terms with
one factor $R_{\mu\nu}$ can be brought in the form with two so do not
produce a pole either. From Table \tref{table1} there 
is no contribution from that 
type of terms to $\ovpi^{(0+1)}_{V,A}$. Then those with
external fields from $\nabla_\mu$ 
with different derivatives necessarily contain
extra factors  $q_\mu q_\nu$ so do not contribute to a possible
 pole at $q^2=0$ and the last type
of terms does not contribute to $\ovpi^{(0+1)}_{V,A}$ as shown above.
This completes the proof.

\setcounter{equation}{0}
\section{Explicit expressions for the barred 
two-point functions}
\rlabel{AppC}

Here we shall give the one-constituent-quark-loop 
expression for the
two-point functions defined in eqs. \rref{9}-\rref{14}
in the presence of current quark masses. These two-point
functions are denoted in the text as the $\ovpi$ ones.
They fulfil the same Ward identities as the full-ones in eqs.
\rref{ward1}-\rref{ward4} changing the current quark masses
there by the constituent quark ones. In addition, they also
satisfy the Ward identities in eqs. \rref{ward5}-\rref{ward6}.
Using these identities one can
see that there are only two independent functions
out of $\ovpi^{(1)}_V$, $\ovpi^{(0)}_V$, $\ovpi^{(1)}_A$,
$\ovpi^{(0)}_A$, $\ovpi_S^M$ $\ovpi_P^M$, $\ovpi_S$
and $\ovpi_P$. We shall take $\ovpi_P^M$ and 
$\ovpi^{(1)}_V + \ovpi^{(0)}_V$ as these functions.
The explicit expressions are

\ba 
\rlabel{piv01}
\left(\ovpi^{(1)}_V + \ovpi ^{(0)}_V \right) 
(Q^2)_{ij} &=& \frac{\dis N_c}{\dis 16 \pi^2}\,
8 \, {\dis \int^1_0} {\rm d}x \,x(1-x) \Gamma(0,x_{ij})\, , \\ 
\rlabel{pimp}
{\overline \Pi}_P^M (Q^2)_{ij} &=& \frac{\dis N_c}{\dis 16 \pi^2}\,
4 {\dis \int^1_0} {\rm d}x (M_i x + M_j (1-x))
\Gamma(0,x_{ij}) \, , 
\ea
where
\ba
x_{ij} &\equiv& \frac{\dis M_i^2 x + M_j^2 (1-x) + Q^2 x(1-x)}
{\dis \Lambda_\chi^2} \,  .
\ea

One can obtain all the others one-loop two-point functions
in function of these two by
using the Ward identities mentioned above.
For instance, for the $\ovpi^{(0)}_V$ one gets

\be
\renewcommand{\arraystretch}{1.5}
\begin{array}{l} \dis
\rlabel{piv0}
{\overline \Pi}^{(0)}_V (Q^2)_{ij} 
= - \frac{\dis \left(M_i - M_j\right)^2}{\dis M_i + M_j} \, 
\frac{\dis \ovpi^M_P (Q^2)_{ij}}{\dis Q^2 (Q^2 + (M_i - M_j)^2)}
\nonumber \\ \hspace*{0.5cm} \times 
\left\{ \left(M_i+M_j\right)^2 + g_A(Q^2)_{ij} m_{ij}^2
(Q^2) \left(1 - \left(\frac{\dis m_i - m_j}{\dis m_i + m_j} 
\right) \left(\frac{\dis M_i + M_j}{\dis M_i - M_j} \right) 
\right) + Q^2 \right\}  \, . \nonumber \\
\end{array}
\renewcommand{\arraystretch}{1}
\ee

\setcounter{equation}{0}
\section{Explicit expression for the one-loop form
factor $\ovpi^+_\mu(p_1,p_2)$}
\rlabel{AppD}

Here we shall give the one-constituent-quark-loop 
expression for the
three-point function $\ovpi^+_\mu (p_1,p_2)$
 defined in eq. \rref{pi+}. We shall give it for $M_i=M_k=M_m$.
The explicit expression is (remember that we have $j=m$),

\ba 
\rlabel{pivpp}
\ovpi^{+\mu}(p_1,p_2) &=& \nonumber \\
&-& \frac{\dis 1}{\dis 2 M_i} \Bigg\{
\ovpi_P^M(-p_1^2)_{ii} \, \nonumber \\
&+&  \frac{\dis p_1\cdot p_2}{\dis p_1^2 p_2^2 -
(p_1 \cdot p_2)^2} \, \left[ p_2^2 
\left(\ovpi^M_P(-p_2^2)_{ii} - 
\ovpi_P^M(-q^2)_{ii}\right) \right. \nonumber \\ 
&+& \left. 
(p_1 \cdot p_2) \left( \ovpi_P^M(-p_1^2)_{ii}
-  \ovpi_P^M(-q^2)_{ii} \right) \right]  \nonumber \\ 
&+& \frac{\dis 2}{\dis M_i} I_3(p_1^2,p_2^2,q^2) 
p_2^2 \left[ 1 + (p_1 \cdot p_2) 
\frac{\dis p_1^2 + (p_1 \cdot p_2)}{\dis p_1^2 p_2^2 - 
(p_1 \cdot p_2 )^2} \right]  \Bigg\} p_1^\mu \nonumber \\
&-& (p_1 \leftrightarrow - p_2) \, . \nonumber \\
\ea  
Where the two-point function $\ovpi^M_P(-p^2)$ was given in 
appendix \ref{AppC} and the function $I_3(p_1^2,p_2^2,q^2)$ is

\ba
I_3(p_1^2,p_2^2,q^2) &=& \frac{\dis N_c}{\dis 16 \pi^2}
2  M_i^2 {\dis \int^1_0} {\rm d}x x {\dis \int^1_0} {\rm d}y
\frac{\dis \Gamma(1, M^2(x,y)/\Lambda_\chi^2)}{\dis
M^2(x,y)} \, 
\ea
with
\ba
M^2(x,y) &\equiv& \nonumber \\ 
&&M_i^2 -p_1^2 (1-x) - p_2^2 x(1-y)
+ (p_1 (1-x) - p_2 x(1-y))^2 \,  .
\nonumber \\
\ea

\newpage
\listoffigures
\end{document}